\documentclass[11pt]{article}

\title{Deep Learning for Constrained Utility Maximisation}

\author{Ashley Davey\footnote{Department of Mathematics, Imperial College, London SW7 2BZ,\, UK. Email: ashley.davey18@imperial.ac.uk}\: and Harry Zheng\footnote{Department of Mathematics, Imperial College, London SW7 2BZ,\, UK. Email: h.zheng@imperial.ac.uk. Supported in part by the EPSRC (UK)  Grant (EP/V008331/1).}}
\date{\today}

\usepackage{graphicx}
\usepackage{float}
\usepackage{framed}
\usepackage{amsmath}
\usepackage{amssymb}
\usepackage{amsfonts}
\usepackage{mathrsfs}
\usepackage{array}
\usepackage{enumerate}
\usepackage{amsthm} 
\usepackage{bbm}
\usepackage{mathtools}
\usepackage{url}
\usepackage{standalone}
\usepackage{comment}
\usepackage{floatpag}
\usepackage{esdiff}
\usepackage{cite}
\usepackage[margin = 1in]{geometry}
\usepackage[ruled]{algorithm2e}
\usepackage{caption}
\usepackage{subcaption}
\usepackage{emptypage}
\usepackage{algorithm2e}
\usepackage{xcolor}

\usepackage{chngcntr}
\numberwithin{equation}{section}

\providecommand{\R}{\mathbb{R}}
\providecommand{\N}{\mathbb{N}}
\providecommand{\p}{\mathbb{P}}

\providecommand{\F}{\mathcal{F}}

\providecommand{\LL}{\mathcal{L}}
\providecommand{\C}{\mathcal{C}}

\providecommand{\Norm}{\mathcal{N}}

\providecommand{\E}{\mathbb{E}}

\providecommand{\ord}{\mathcal{O}}
\providecommand{\ind}{\mathbbm{1}}
\providecommand{\half}{\ensuremath{\frac{1}{2}}}
\providecommand{\tr}{\text{tr}}
\newcommand*{\defeq}{\mathrel{\vcenter{\baselineskip0.5ex \lineskiplimit0pt
                     \hbox{\scriptsize.}\hbox{\scriptsize.}}}%
                     =}
\newcommand*{\eqdef}{=\mathrel{\vcenter{\baselineskip0.5ex \lineskiplimit0pt
                     \hbox{\scriptsize.}\hbox{\scriptsize.}}}%
                     }

\newtheorem{theorem}{Theorem}[section]

\newtheorem{remark}[theorem]{Remark}

\theoremstyle{definition}

\newtheorem{example}[theorem]{Example}

\begin{document}
\date{}% working version
\maketitle

\begin{abstract}
This paper proposes two algorithms for solving stochastic control problems with deep  learning, with a focus on the utility maximisation problem. The first algorithm solves Markovian problems via the Hamilton Jacobi Bellman (HJB) equation. We solve this highly nonlinear partial differential equation (PDE) with a second order backward stochastic differential equation (2BSDE) formulation. The convex structure of the problem allows us to describe a dual problem that can either verify the original primal approach or bypass some of the complexity. The second algorithm utilises the full power of the duality method to solve non-Markovian problems, which are often beyond the scope of stochastic control solvers in the existing literature. We solve an adjoint BSDE that satisfies the dual optimality conditions. We apply these algorithms to problems with power, log and non-HARA utilities in the Black-Scholes, the Heston stochastic volatility, and path dependent volatility models. Numerical experiments show highly accurate results with low computational cost, supporting our proposed algorithms.
\end{abstract}


\section{Introduction}

In this paper we propose algorithms that combine modern machine learning practises with theoretical stochastic control principles to solve a range of control problems with high accuracy, scalability with dimension, and low computational cost. There is a natural crossover between machine learning and stochastic control problems, as they both involve searching for features within a data set. For stochastic control the data set is a state process that is influenced by a controlling process that can be chosen with the aim to optimise some objective function. Often the optimal control is a function of the underlying state process, a so-called Markovian control. This is very similar to deep  learning, where the control is chosen as the output of a neural network which takes the dataset as input, and some loss function is minimised. It is in this area of crossover that this paper resides. We propose deep learning methods to solve the utility maximisation problem, as an application of a more general stochastic control solver, in a wide range of markets with arbitrary convex control constraints. Dynamic portfolio optimisation has been extensively studied within the field of mathematical finance, see \cite{pham2009continuous, karatzas1998methods} for exposition.

A typical drawback of numerical methods for stochastic control problems is that their complexity increases dramatically with increases in the state and control dimensions. Often some grid method or exhaustive search of the underlying state space is required to find optimal trajectories, which is computationally inefficient for high dimensional state spaces. It is claimed that machine learning methods provide a solution for this dimensionality problem \cite{grohs2018proof}. Numerical results show efficient solving of PDEs with the state space dimension in the thousands \cite{beck2019deep, beck2017machine, E2017}. These PDE solvers form the basis of our generic stochastic control problem solver, which is the first algorithm presented in this paper. The problems are transformed into forward-backward stochastic differential equations (FBSDEs) using the Feynman-Kac formula and the Stochastic Maximum Principle (SMP). The PDE that we solve is the Hamilton-Jacobi-Bellman (HJB) equation, a highly nonlinear PDE that is often impossible to solve analytically, or even  numerically   if the dimensionality is high. This paper introduces a control variable into the PDE, that is constrained to some convex set. This drastically increases the complexity and takes the problem out of the scope of \cite{E2017,beck2017machine} unless the control can be found analytically. We show how machine learning can overcome this difficulty and produce highly accurate  value approximations with comparatively low runtime.

The FBSDE based methods developed in this paper are direct extension of the first and second order deep BSDE solvers introduced in \cite{beck2017machine, E2017}. In these papers, a system of FBSDEs is solved by simulating all processes in the forward direction, driven by a neural network process which is optimised to ensure the terminal condition holds. The extension offered by this paper is an additional driving neural network process, representing a control process. The convenience of adapting this BSDE method to stochastic control problems is that the control process can be optimised using the value function approximations implied by the 2BSDE solver, via the Hamiltonian of the system. In the 2BSDE solver of \cite{beck2017machine}, there are two unknown process that must be found with neural networks. This suggests that our method has three unknowns, but we can use the structure of the value problem to derive one of these processes and therefore introduce control without increasing the number of unknown processes. 

There are some papers that describe generic algorithms for solving stochastic control problems in the literature. \cite{han2016deep} directly maximises the gains function at the terminal time with machine learning and is applicable to many problems. However, the computation costs may be high, as one introduces a discretisation with many time steps  or a high dimensional control space, and it only outputs the value at time 0. \cite{hure2018deep1}  finds a local approximation of the value function  with machine learning.  However, it  only holds for Markovian problems and  the iterative nature of this value function approximation may lead to error propagation.   
Instead of directly optimising the value function, the first method presented in this paper maximises the Hamiltonian associated with the system. There are two key benefits of doing this. Firstly, it reconciles the potential errors of \cite{han2016deep, hure2018deep1} where the algorithm gets stuck at a local optimiser. To find the value function (given a control) we ensure that the terminal condition of the HJB equation holds, which involves minimising the least squares error of the associated processes at the terminal time. The global minimum of this error is zero by construction, ensuring that we can directly assess the quality of our approximation for any generic control problem. Secondly,   the control is optimised using the Hamiltonian of the system, which is a function of only what is known at one time. This optimisation is faster than back propagating through a discretisation of the processes to compute the derivatives as in \cite{han2016deep}. This is even more important when the control space is high dimensional, which is normally true in portfolio optimisation problems.  To implement this Hamiltonian maximisation procedure, one needs to know derivatives of the value function which can be computed accurately with deep learning based on the 2BSDE method in \cite{beck2017machine}. 

There are a number of papers in the literature that utilise the deep BSDE methods of \cite{E2017, beck2017machine}. For example,  \cite{pereira2019neural, wang2019deep, exarchos2018stochastic} first find an explicit formula for the optimal control in terms of the value function and its derivatives and then substitute this control  into the HJB equation and solve the resulting PDE with the deep 2BSDE method. However, these algorithms cannot be applied in full generality like our method.  \cite{pereira2019neural, exarchos2018stochastic} are reliant on the diffusion of the state process being independent of the control, which leads to a semi-linear PDE and a representation with a first order BSDE. This framework is  inappropriate for portfolio selection  problems as the control naturally appears in the diffusion term. \cite{wang2019deep} alleviates this restriction by deriving an explicit formula for the optimal control that is required to be unconstrained. Our approach is   more general in the sense that  it does not require any analytical solutions. In the literature these other algorithms are called ``deep 2BSDE controllers'' since the BSDE solutions are used to form the control. Our algorithm is called  the deep controlled 2BSDE method as each control and state pair defines its own 2BSDE which we try to solve. A primal-dual method is introduced in \cite{henry2017deep} for solving semi-linear PDEs. The deep 2BSDE method is used to find a lower bound for the solution. An upper bound is found by solving a maximisation control problem on the path space, that is derived using the Legendre transform. The approach in \cite{henry2017deep} is not directly related to our notion of convex duality using martingale methods. There are also some papers on applications of deep neural networks in mathematical finance, including optimal stopping \cite{becker2019}, asset pricing with transaction cost \cite{gonon2020}, and optimal transport \cite{eckstein2019}. 

The utility maximisation problem involves maximising concave functions, so we can naturally apply convex duality methods. Duality is a powerful tool that can be used in complete or incomplete markets with a range of convex constraints on the control process. The unique aspect of our work is the combination of deep learning methodology with convex duality. The benefits of introducing duality are threefold. Firstly, the dual problem is a minimisation problem, and solving it independently of the primal problem naturally leads to an upper bound of the value function. Sometimes the dual problem is easier to solve, since primal control constraints are characterised by penalties in the dual state process, leading the dual problem to be potentially unconstrained. Secondly, the BSDEs associated to the dual problem can be related to that of the primal problem, giving us the ability to compare our numerical results and check their accuracy. Again, as per the Hamiltonian maximisation, this benefits us since we know the error between the dual and primal processes should be zero, which allows us to easily measure the quality of our results. We may even skip one of the problems if they are difficult to solve, as solving one problem automatically gives us the solution to the other. Thirdly, inherent in the dual formulation are certain optimality conditions that are derived from the SMP. These conditions can be used separately from the HJB equation to find the optimal control. We use these conditions as the basis for our second algorithm, called the deep SMP method, which is valid for non-Markovian control problems with convex structure. This greatly increases the range of utility maximisation  problems that can be solved using the methods of this paper. In a Markovian setting we can compare the two algorithms presented in the paper. It should be noted here that results for dynamical programming principles, under which our 2BSDE scheme is formulated, exist in the literature for non-Markovian problems. Non-Markovian optimal stopping problems and their 2BSDE counterparts are studied in \cite{karoui2013capacities, possamai2018stochastic}. The problem is reformulated into an optimisation in the space of probability measures, and the canonical probability space is expanded to allow a dynamical programming principle and a characterising path dependent PDE to be established, see \cite{zhang2017backward} for an exposition. However, it is not clear how a machine learning based solver could be used in this abstract setting.

The convex duality method works for complete and incomplete markets. A common incomplete market model is the Heston stochastic volatility model in which the underlying variance of the stock follows some mean-reversion SDE with its own source of randomness. This model may be seen as a 2-dimensional stock market, where only one asset is traded. This framework allows us to use our deep learning method to solve it. We compare our results to \cite{ma2020dual} in which 
tight bounds of the value function are derived for an unconstrained optimisation problem. \cite{ma2020dual} is reliant on a convenient guess of the dual control, so is not applicable to other methods where such a guess is not available, unlike our method. Another multidimensional model we consider is a path-dependent  volatility model.  There are several recent papers on the topic  in mathematical finance with machine learning, for example \cite{buehler2019deep,liu2019pricing, de2018machine}. These papers include model calibration, hedging and pricing, and are not directly related to our work. 

The main contributions of this paper are the following. In the deep controlled 2BSDE algorithm, we present a Markovian stochastic control problem solver that combines the HJB equation with the deep 2BSDE method, but without the need for an analytical formula for the optimal control. We apply this algorithm to both the primal and dual problems which give us tight bounds on the value function. In the deep SMP method, we exploit the dual formulation of the utility maximisation problem to create a novel algorithm for non-Markovian optimisation problems using deep  learning. Instead of dealing with the gains function, we solve the adjoint BSDE and use machine learning to satisfy the dual optimality conditions. To the best of our knowledge, we are the first to use this methodology in the literature.

The remainder of this paper is outlined as follows. Section 2 formulates the utility maximisation problem and defines the dual problem. Section 3 describes the general algorithm for solving Markovian stochastic control problems using deep learning and the HJB equation, and applies the algorithm to portfolio optimisation with various utilities. Section 4 introduces an algorithm for solving non-Markovian utility maximisation problems using the SMP and applies this algorithm to problems where the volatility process is not deterministic, and even path-dependent. Section 5 discusses convergence of the deep controlled 2BSDE algorithm based on empirical evidence, and compares to other algorithms. Section 6 concludes the paper.

\section{The Utility Maximisation Problem}
In this section we describe  the standard portfolio optimisation problem and its dual problem. 
Let $W=(W(t))_{0 \leq t \leq T}$ be a standard $m$-dimensional Brownian motion on the natural filtered probability space $(\Omega, \F,(\F_t)_{0 \leq t \leq T},\p)$ augmented with all $\p$-null sets, and $T$ a fixed finite horizon. 
Consider an $m+1$ dimensional stock market, consisting of a risk-free bond with interest rate $r(t)$, and a collection $S$ of risky stocks defined by
\begin{align}
dS(t) = \text{diag}(S(t))\left(\mu(t) dt + \sigma(t) dW(t)\right), \label{market}
\end{align}
where $\text{diag}(S(t))$ is an $m \times m$ matrix with diagonal elements from $S(t)$ and all other elements zero, and $r$, $\mu$ and $\sigma$ are uniformly bounded, progressively measurable processes  valued in $\R$, $\R^m$ and $\R^{m \times m}$, respectively, with $\mu(t)$ and $\sigma(t)$ representing the drifts and volatilities of the $m$ stock prices at time $t$, respectively. The matrix $\sigma(t)$ is invertible and $\sigma^{-1}$ is uniformly bounded. An agent invests a proportion $\pi(t)$ of the wealth $X(t)$ in the risky stocks  and the rest in the risk-free bond at time $t \in [0,T]$.  The 1-dimensional wealth process $X$ evolves as
\begin{align}
dX(t) = X(t)\left(r(t) + \pi(t)^\top \left(\mu(t) - r(t)\ind_m\right)\right) dt + X(t)\pi(t)^\top \sigma(t) dW(t)
\label{wealth}
\end{align}
with $X(0) = x_0$,  where $\pi(t)^\top$ is the transpose of $\pi(t)$, and $\ind_m \in \R^m$ is a vector of ones.  We call $\pi$ an admissible control if  $\pi$ is a progressively measurable process valued in $K \subset \R^m$, a non-empty closed convex set, almost surely for almost every $t \in [0,T]$, and there exists a unique strong solution $X$ to the SDE (\ref{wealth}). The set of all admissible controls $\pi$ is denoted by ${\cal A}$.
The investor wishes to maximise the expected utility of the terminal wealth, that is, to find
\begin{align*}
V \defeq \sup_{\pi \in {\cal A}} \E\left[U(X(T))\right],
\end{align*}
where $U \colon [0,\infty) \to \R$ is a utility function  that is twice continuously differentiable, strictly increasing, strictly concave and satisfies the Inada conditions
$U'(0) = \infty$ and $U'(\infty) = 0$.
We say that state and control processes $\hat{X},\, \hat{\pi}$ are optimal if $V = \E[U(\hat{X}(T))]$.
%\begin{remark}
%The set $K$ to which we restrict our strategies can often be inferred from real-world market data. We are normally always allowed to buy arbitrary units of any stock, that is, $\R_+^m \subseteq K$. The remainder of the space $K$ is the set of non-negative linear combinations of the trades given by bid-ask matrices, forming a convex cone, known as the ``solvency cone" \cite{hamel2010duality}. However, there may also be constraints on trading strategies not from the market such as maximum exposure, which would make $K$ the intersection of the solvency cone and some ball $B \subset \R^m$ centred at the origin. Other complex constraints are possible, but the assumption of convexity should always hold. If an investor can implement two portfolios, they should be able to diversify and implement any convex combination of them. 
%\end{remark}

We next construct the dual problem, see  \cite[Section~2]{li2018dynamic} for more details. 
The basis for duality is the Legendre-Fenchel transformation $\tilde{U} \colon (0,\infty) \to \R$ of the utility function $U$, defined by
\[\tilde{U}(y) = \sup_{x > 0}\left\{ U(x) - xy \right\}, \quad y\geq 0.\]
The concavity and differentiability of $U$ ensure that $\tilde{U}$ is well defined for all $y\geq 0$ and  is twice continuously differentiable, strictly decreasing and strictly convex on $(0,\infty)$.
The $\R$-valued dual process $Y$ is defined by
\begin{align}
dY(t) & = -Y(t) \big(r(t) + \delta_K(v(t))\big) dt - Y(t) \left(\theta(t) +\sigma(t)^{-1}v(t)\right)^\top dW(t) \label{Yupdate}
\end{align}
with $Y(0)=y$, where $y$ is non-negative,  $\theta(t) = \sigma(t)^{-1}\left(\mu(t) - r(t)\ind_m\right)$ is the market price of risk, $\delta_K$ is  the support function  of the set $-K$, defined by 
\[\delta_K(z) \defeq \sup_{\pi \in K} \left\{ -\pi^\top z\right\}, \quad z \in \R^m, \]
and $v$ is a $\R^m$-valued  dual control process which is progressively measurable and ensures there exists a unique strong solution $Y$ to the SDE (\ref{Yupdate}). 

Since  $XY$ is a super-martingale for any choice of $\pi,\, v$ and $y$, we have 
 the following weak duality relation \cite[p.~5]{li2018dynamic} 
\begin{align}
V \leq \tilde{V}(x_0) \defeq  \inf_{y > 0} \left\{ \inf_{v \in \R^m} \left\{\E\left[\tilde{U}(Y (T)) \,\middle|\, Y (0) = y \right]\right\} + x_0 y  \right\}. 
\label{duality}
\end{align}
Solving the dual problem would give us an upper bound of the value function, that can be combined with a lower bound by the primal problem to produce a confidence interval for the value function. If we have an equality in (\ref{duality}), then we can find the value function by alternatively solving the dual problem. 

\begin{comment}
\begin{remark} \label{constraints}
If $K = B(0,R)$ is a ball of radius $R>0$ centred at the origin, then $\delta_K(z) = R |z|$. If $K$ is a closed convex cone, then 
$\delta_K(z) =0$ if $z \in \tilde{K}$ and $\delta_K(z) =\infty$ if $z \notin \tilde{K}$,  
where $\tilde{K}$ is the positive polar cone of $K$, defined by $\tilde{K} \defeq \{z \in \R^m \colon \pi^\top z \geq 0 \text{ for all } \pi \in K\}$, which greatly simplifies the dual problem as the drift is independent of the dual control $v$ if $v(t)\in \tilde{K}$. In particular, if $K = \R^m$,  then  $\tilde{K} = \{0\}$, which forces the dual control $v=0$. 
For general closed convex set $K$, the function $\delta_K$ is well defined but may be difficult to compute and cannot be exactly implemented numerically due to the infinite penalty. Instead, so-called `soft' constraints can be used, where a suitably large penalty function gives us sufficient confidence that our control process lies in the required space. Sometimes it is easier to implement a `hard' constraint, where the optimal control is projected onto the space that makes this function finite before optimisation as in the cone case. With hard constraints, we can have 100\% confidence that the process lies in the required space by construction.
\end{remark}
\end{comment}

The method  of solving both the primal and dual problems depend on the nature of the coefficients in (\ref{wealth}). If $r$, $\mu$ and $\sigma$ are deterministic, then the problem is Markovian. This case is treated in Section \ref{Markovian}, where we solve the HJB equation numerically using deep learning, and the control is optimised using the Hamiltonian. If the coefficients are not deterministic then the problem may not be Markovian. We treat this more general case in Section \ref{Non-Markovian}, where the adjoint BSDE is solved using deep learning, and the optimality conditions derived in \cite{li2018dynamic} are used to optimise the control.

\section{The Deep Controlled 2BSDE Method for the Markovian Case} \label{Markovian}

The deep learning approaches for both the dual and primal Markovian problems are similar, so we present the algorithm in a more general form that covers both. We first describe the general control problem, then  use the machine learning method to find the value function and optimal control. For a thorough exposition of the theory covered in this section, we direct the reader to \cite[Chapter~3]{pham2009continuous}. We present the algorithm for a state dimension $d \geq 1$, as this is required in proceeding sections. For example, when we consider the Heston model with a stochastic volatility process denoted by $\sigma \in \R^m$, we can make the problem Markovian by introducing new (1 + $m$) dimensional state processes $(X, \sigma)$ and $(Y, \sigma)$.

Let $(\Omega, \F , (\F_t)_{0 \leq t \leq T} , \p)$ be a filtered probability space with a fixed finite horizon $T$. Consider the $\F_t$-adapted controlled diffusion process $\left(\mathcal{X}(t)\right)_{0\leq t \leq T}$ on $\R^d$, $d \in \N$, defined by
\begin{align}
d\mathcal{X}(t) = b\left(t,\mathcal{X}(t),\alpha(t)\right)dt + \Sigma\left(t,\mathcal{X}(t),\alpha(t)\right)dW(t) \label{update}
\end{align}
with $X(0) = x_0$, where $W$ is an $\R^n$-valued standard Brownian motion for some $n \in \N$ and $\alpha = \left(\alpha(t)\right)_{0\leq t \leq T}$ is a progressively measurable process, valued in $A \subset \R^m$ for $m \in \N$. It is assumed that the drift and diffusion functions $b \colon [0,T] \times \R^d \times A \to \R$ and $\Sigma \colon [0,T] \times \R^d \times A \to \R^{d \times n}$ are deterministic and measurable, such that (\ref{update}) admits a strong solution. Here $\mathcal{X}$ represents the  state process and $\alpha$ the control process. We consider only admissible controls that satisfy the integrability condition
\[ \E\left[\int_0^T |b(t, x, \alpha(t))|^2 + |\Sigma(t, x, \alpha(t))|^2dt\right] < \infty \] 
where $x$ is some element of the support of $\mathcal{X}$. This control is chosen to optimise the state process via maximising  a reward that is defined using a terminal gain function $g \colon \R^d \to \R$, which is a measurable function, either lower bounded or satisfying some quadratic growth condition. We wish to find the value function
\begin{align}
u(t,x) \defeq \sup_{\alpha \in {\cal A}} \E \left[ g\left(\mathcal{X}(T)\right) \,\middle|\, \mathcal{X}(t) = x\right]. \label{value}
\end{align}
\begin{remark} \label{terminal} The utility function $U$ is not defined on the whole space $\R$, but only $(0,\infty]$. We take $g(x)$ to be $U(x)$ when $x > 0$ and 0 otherwise, ensuring that quadratic growth or bounded conditions are satisfied (if they are satisfied for $U$). It is irrelevant what value is chosen when $x < 0$, as one can show that the state process $X$ defined in (\ref{wealth}) satisfies $X(T) > 0$ if and only if $X(0) > 0$ almost surely. We can similarly consider $\tilde{U}$ and $Y$.
\end{remark}
Suppose that $u \in \C^{1 ,3}$. The HJB equation tells us that if $u$ has at most quadratic growth in $x$ then it solves the following PDE
\begin{align} \label{HJB}
\frac{\partial}{\partial t} u(t,x)  = - \sup_{a \in A} F(t,x,a, D_x u(t,x), D_{xx} u(t,x)), \quad (t,x) \in [0,T) \times \R^d
\end{align}
with the terminal condition $u(T,x)  = g(x)$ for $x \in \R^d$, 
where the Hamiltonian $F$ is defined by
\begin{align}
F(t,x,a,z,\gamma) = b(t,x,a)^\top z + \half \tr \left(\Sigma(t,x,a)\Sigma(t,x,a)^\top \gamma \right). 
\label{ham}
\end{align}

We can now describe the value function and its derivatives using a system of FBSDEs. This  involves the generalised Hamiltonian $H \colon  [0,T) \times \R^d  \times A \times \R^{d} \times \R^{d \times n} \to \R.$ defined by
\begin{align}
H(t,x,a,z,q) = b(t,x,a)^\top z +  \tr \left(\Sigma(t,x,a)^\top q \right).
\label{gen_ham} 
\end{align}

\begin{theorem} Suppose that $u \in \C^{1 ,3}([0,T) \times \R^d) \cap \C^0([0,T] \times \R^d)$, and there exists an optimal control $\alpha \in {\cal A}$ with associated controlled diffusion $X$, defined by (\ref{update}). Then there exist continuous processes $(V, Z, \Gamma)$, valued in $\R$, $\R^d$, and  $\R^{d \times d}$ respectively, solving the following 2BSDE:
\begin{align}
\begin{split}
dV(t) &= Z(t)^\top \Sigma(t, \mathcal{X}(t),\alpha(t)) dW(t) \\
dZ(t) & = - D_x H(t,\mathcal{X}(t), \alpha(t), Z(t), \Gamma(t) \Sigma(t,\mathcal{X}(t), \alpha(t))) dt + \Gamma(t) \Sigma(t,\mathcal{X}(t), \alpha(t)) dW(t)  
\end{split} \label{2BSDE}
\end{align}
with the terminal conditions $V(T)  = g(\mathcal{X}(T))$ and $Z(T)  = D_xg(\mathcal{X}(T))$, 
where $H$ is defined by (\ref{gen_ham}) and the control $\alpha$ satisfies
\begin{align}
F\left( t, \mathcal{X}(t),\alpha(t),Z(t),\Gamma(t)\right) = \sup_{a \in A} F\left(t,\mathcal{X}(t),a,Z(t),\Gamma(t)\right),  \label{maxHam}
\end{align}
and $F$ is defined by (\ref{ham}).
\label{2BSDE_thm}
\end{theorem}

A proof  is provided in the appendix. The derivative term in the dynamics of $Z$ can be easily computed by the modern automatic differentiation method if it is unavailable analytically. This theorem is presented for a maximisation problem but easily translates to the dual problem, which is a minimisation problem, by replacing supremums with infimums.

\subsection{The Deep Controlled 2BSDE Method} \label{ML}

Now we have set up the new problem (\ref{2BSDE})-(\ref{maxHam}), we look to solve it with machine learning techniques in the spirit of \cite{beck2017machine} (the algorithm presented therein is named the ``Deep 2BSDE'' method, thus inspiring our adapted Deep Controlled 2BSDE method). The trick here is to simulate all processes in the forward direction, introducing new variables $v^0_0 \in \R$ and $z^0_0 \in \R^d$, then moving forward through a discretisation $(t_i)_{i=0}^N$ of $[0,T]$. We furthermore set $\Gamma$ to be a neural network that at each time $t_i$ takes the $i$th state position, denoted by $X_i$, as input since we are approximating $D_{xx} u(t,X(t))$ with this process. 
The control process $\alpha$ is also a neural network, taking in the same state $X_i$ as input at time step $t_i$, which means that we are searching for a Markovian control.

Before describing the algorithms used to solve this problem in detail, we first look at what we mean by a neural network. We give a brief outline here, see for example \cite{anthony2009neural} for an exposition. Suppose we wish to approximate a function $ \phi \colon \R^p \to \R^q$ for some $p,q \in \N$. We use the following algorithm to construct the approximating function. This neural network  has a predetermined amount of `layers' $L \in \N$ and `hidden nodes' $\ell \in \N$. We call these the hyper-parameters of the network in the sequel. 

Let $X \in \R^p$ be our input. For $i = 1, \ldots, L$ define the linear function $f_i$ by
\[f_i(x) = A_i x + b_i \]
We  have $A_1 \in \R^{\ell \times p}$, $A_j \in \R^{\ell \times \ell}$ for $j = 2, \ldots, L-1$ and  $A_L \in \R^{q\times \ell }$ for the matrices. For the vectors we  have $b_j \in \R^\ell$ for $j = 1, \ldots, L-1$ and $b_L \in \R^q$. These form the unknowns of our function, and are what we need to choose to correctly approximate our function. We combine these into a set $\theta = (A_i, b_i)_{i=1}^L$ of parameters. Informally speaking, the number of hidden nodes $\ell$ corresponds to us `expanding' the input into a higher dimensional space. One could think of each node as a piece of information we obtain from the input. The more nodes we have the more information we get, hence the better we can approximate the function. We wish to compose these linear functions in series, but we need a nonlinear function in between them. Let $h \colon \R \to \R$ be such a function that we apply element wise to the output of each linear function. The approximating function is therefore
\[N_\theta(X) \defeq f_L \circ h \circ f_{L-1} \circ \ldots \circ f_{2} \circ h \circ f_1 (X) \]
In our setting we choose $h(x)= \max(x,0)$, called a rectified linear unit (ReLU) activation function. The choice of this function for this scheme is rather arbitrary, as there seems to be no clear preferred function here. A comparison of popular activation functions in provided in Section \ref{method}. This non-polynomial mapping is essential in ensuring the density of these networks within the space of continuous functions, as is shown in \cite[Theorem~1]{leshno1993multilayer} which formalises the informal argument above about information into a density argument about the space of neural networks. Further more, if $\phi$ is Lipschitz then we can bound the approximation error in terms of the hyper-parameters of the network \cite[Proposition~6]{bach2017breaking}. Note that no nonlinear function is applied at the final step, else we could only approximate non-negative functions! The parameter vector $\theta \in \R^\rho$, where $\rho = \rho(p,q)$ is defined below, is a parametrisation of the $A_i$ and $b_i$ matrices and is optimised at the training stage of our algorithm.
We made no assumptions on the regularity of $\phi$, but our approximation is almost everywhere differentiable. In practise this is sufficient for gradient descent algorithms, in which derivatives of this network with respect to its parameters must be taken. We use the \texttt{Python} package \texttt{Tensorflow} to implement this neural network procedure.
This methodology reduces an optimisation over an infinite dimensional space to a finite dimensional one. However, the dimension can still be high. For this neural network, the parameter space dimension is
\begin{align}
\rho(p,q) = (L-1)\ell^2 + \ell(L + p+ q) + q,\label{NN_dim}
\end{align}
a quadratic in the number of hidden nodes, and linear in the number of layers and the input dimension. 
Associated to each neural network $N_\theta$ is a loss function $\LL(\theta)$ which is a function of the parameters forming the network. Finding the parameters that minimise this function equates to optimising the network itself. In our implementation, we use the ADAM algorithm \cite{kingma2014adam}, which is an adaptation of stochastic gradient descent, implemented in \texttt{Tensorflow}. We furthermore use mini batches, which means the number of simulations we use to calculate the loss function is small, and batch normalisation, meaning that we scale our input to the neural network \cite{ioffe2015batch}.

\begin{remark} In this algorithm we propose a control $\alpha$ of the form $\alpha(t) = N_\theta(\mathcal{X}(t))$. One may then ask if this control is indeed admissible. It can be shown that a neural network is a Lipschitz function (though a global Lipschitz constant for every neural network in the training process may not be established), and therefore the required integrability properties of the state process and gains function of the control problem are satisfied.
\end{remark}

Now we describe the algorithm to solve the control problem. We introduce parameters $\theta^0_i \in \R^{\rho (d,m)}$ and $\lambda^0_i \in \R^{\rho(d,d^2)}$ then use the following Euler-Maruyama Scheme. 
Set $\mathcal{X}_0 = x_0,\, V_0 = v_0^0$ and $Z_0 = z_0^0$ then for $i = 0, \ldots N-1$ let $\alpha_{i} = N_{\theta_i^0}(\mathcal{X}_{i})$, $\Gamma_{i} = N_{\lambda_i^0}(\mathcal{X}_{i})$ and
\begin{align}
\begin{split}
\mathcal{X}_{i+1} & = \mathcal{X}_i + (t_{i+1}-t_i)b(t_i, \mathcal{X}_i,\alpha_i) + \Sigma(t_i, \mathcal{X}_i,\alpha_i)dW_i \\
V_{i+1} & = V_i +  Z_i^\top \Sigma(t_i, \mathcal{X}_i,\alpha_i) dW_i \\
Z_{i+1} & = Z_i - (t_{i+1}-t_i)D_x H(t_i,\mathcal{X}_i, \alpha_i, Z_i, \Gamma_i \Sigma(t_i, \mathcal{X}_i, \alpha_i)) + \Gamma_i \Sigma(t_i, \mathcal{X}_i,\alpha_i) dW_i,
\end{split} \label{2BSDE_disc}
\end{align}
where $dW_i$ is a multivariate normal $\Norm_n(0,(t_{i+1}-t_i)\ind_n)$ random variable. We have used minimal notation for these processes for clarity, but one should remember that these explicitly depend on the choice of parameters $\lambda_i^0$, $\theta_i^0$ and the initial points $v^0_0$, $z_0^0$. Our algorithm involves using this scheme multiple times, but with different parameters.

{
\begin{remark} \label{one_net}
We use a different neural network at every time step, as opposed to a single network that takes in state and time as inputs. While this increases the number of parameters (assuming all networks use the same number of layers and hidden nodes), it also decreases the run time. This is because in the back propagation step, extra time is required to take derivatives of the network parameters, since they are involved in every time step. Even when using a high number of time steps, the increase in time outweighs the penalty of having more networks which leads to more memory usage. See Section \ref{parameter} for a numerical example.
\end{remark}
}
Once we have repeated this iteration $N-1$ times, we arrive at time $t_N = T$ but realise that we have not satisfied the terminal conditions. We  want to choose the neural networks and initial start points to reduce the expected error, which amounts to minimising the the loss function
\begin{align}
\LL_1(\Theta^0_{\text{BSDE}}) \defeq \E \left[\left| V_N - g(\mathcal{X}_N)\right|^2 + \gamma \left|Z_N - D_x g(\mathcal{X}_N)\right|^2\right],  \label{L_1}
\end{align}
where $\Theta^0_{\text{BSDE}} \in \R^{1 + d + \rho(d,d^2)}$ is a vector representation of $(v^0_0, z^0_0 ) \cup (\lambda^0_i)_{i=0}^{N-1}$. This is referred to as the BSDE loss function in the sequel. In practise, we evaluate the sample average of this loss function. The coefficient $\gamma >0$ is some tuning parameter that can be chosen. In our case, we choose $\gamma = 0.5$.

In addition, for each time step the arbitrary control choice $\alpha$ is not optimal, so we must also maximise the loss functions
\begin{align}
\LL_2(\theta^0_i,i) \defeq \E\left[ F( t_i, \mathcal{X}_i,\alpha_i, Z_i, \Gamma_i)\right] \label{L_2}
\end{align}
for $i = 0, \ldots , N-1$. This is referred to as the control loss function in the sequel. Again, we evaluate a sample average of this loss function. We proceed to optimise these loss functions in turn, starting with our arbitrary 0-superscript parameters. Now suppose we have just finished step $j \in \N$.

The first sub-step is to improve the terminal condition loss. We generate $V_N, Z_N$ and $\mathcal{X}_N$ using $\Theta^j_\text{BSDE}$ and $\theta^j_i$ for $i= 1 , \ldots , N-1$, which are carried forward from the previous step. We then make an improvement using one step of the ADAM algorithm, against the loss function $\LL_1$, resulting in the updated parameters $\Theta^{j+1}_\text{BSDE}$. We also keep track of the moment estimates that are present in the algorithm as in \cite{kingma2014adam}. 

The second half of the iteration step is to improve the control process. We can first simulate the processes $\alpha$ and $\mathcal{X}$ using our previous parameter set $\theta^{j} = (\theta^{j}_i)_{i=0}^{N-1}$, then simulate $Z$ and $\Gamma$ using the updated parameters $\Theta^{j+1}_\text{BSDE}$, and finally make the following improvement using one step of the ADAM algorithm. For each $i= 0 ,  \ldots , N-1$ we update $\theta^j_i$ using the ADAM algorithm against $-\LL_2(\cdot,i)$, resulting in $\theta^{j+1}_i$. We use a negative here since ADAM is a minimisation algorithm, but we wish to maximise the Hamiltonian $F$ at each time step. We track the moment estimates and move to step $j+2$ and repeat this process. We do this for each path of a batch of simulated paths of size $B \in \N$.

These two steps can be repeated until the new parameters are sufficiently close to the old parameters after an update, at which point we deem the algorithm to have converged. We find that making small increments at each step, ensuring that the learning rate required for the ADAM algorithm is smaller for the control step than the BSDE step, we have good convergence results. 

\begin{remark}
At each time step we build 2 neural networks, which have 4 layers, with the two hidden layers consisting of $d+10$ nodes. The parameters are optimised using the ADAM algorithm, using a batch size of 64 Brownian paths. We choose to initialise each parameter near 0, though a prior guess may be used, and the rate of convergence is naturally much higher when the initial guess is closer to the true value. Substituting $p = d$, $L = 4$ and $l = d + 10$ into (\ref{NN_dim}) leads to a parameter set of size $\rho(d, q) = 4d^2 + 74d + 340 + q(d + 11)$, where $q$ is the output dimension of the network. The number of neural network parameters is therefore at most cubic in the state dimension and linear in the control dimension.
\end{remark}
{
\begin{remark} \label{one_loss}
We use a different loss function for the BSDE and control networks in the algorithm. Another option would be combining these loss functions into one single loss function that both networks try to minimise. However, this would be slower as the back propagation step would take longer, particularly for the control network. It may also lead to false solutions, for example minimising $\LL_1$ by making both $V_N$ and $g(\mathcal{X}_N)$ close to zero. See Section \ref{parameter} for a numerical example.
\end{remark}
}
Algorithm \ref{primal_DC2BSDE} in the appendix outlines this scheme explicitly. When we use the notation ADAM(), we refer to one step of the ADAM algorithm outlined in \cite[Algorithm~1]{kingma2014adam}. The output is given as the value at time 0, but a byproduct of this scheme is pathwise evaluations of the value function and its derivatives, as well as the optimal control.

\subsection{Extension of the Algorithm to the Dual Problem}
The dual problem (\ref{duality}) has a similar structure, so we can adapt this method to solve it. We  stay in a multidimensional setting, but this one dimensional parameter is just one component of the initial state. We  define a variable $y^0_0$ with which the state process (which for the dual problem we denote by $\mathcal{Y}_i$ for $i=0,\ldots ,N$) begins, as opposed to starting at $x_0$. We only optimise one element of the initial dual state, but keep  a multi dimensional state as $y^0_0 = (y,\tilde{y}) \in \R^d$, where $\tilde{y} \in \R^{d-1}$ is known. For example, when we consider the stochastic volatility model $\tilde{y}$ is the volatility process at time 0, which is known.
Suppose $\mathcal{Y}$ is driven by a control $\beta$, a progressively measurable process, valued in $B \subset \R^m$, and has (deterministic and measurable) drift and diffusion functions $\tilde{b} \colon [0,T] \times \R^d \times A \to \R$ and $\tilde{\Sigma} \colon [0,T] \times \R^d \times A \to \R^{d \times n}$ respectively. Then, in analogue to Theorem \ref{2BSDE_thm}, we have solutions $(V, Z, \Gamma)$ to the 2BSDE
\begin{align}
\begin{split}
d\mathcal{Y}(t) & = \tilde{b}(t, \mathcal{Y}(t), \beta(t)) dt + \tilde{\Sigma}(t, \mathcal{Y}(t), \beta(t)) dW(t) \\
dV(t) &= Z(t)^\top \tilde{\Sigma}(t, \mathcal{Y}(t),\beta(t)) dW(t) \\
dZ(t) & = - D_x \tilde{H}(t,\mathcal{Y}(t), \beta(t), Z(t), \Gamma(t) \Sigma(t,\mathcal{Y}(t), \beta(t))) dt + \Gamma(t) \tilde{\Sigma}(t,\mathcal{Y}(t), \beta(t)) dW(t)  
\end{split} \label{2BSDE_dual}
\end{align}
with the terminal conditions $V(T)  = \tilde{g}(\mathcal{Y}(T))$ and $Z(T)  = D_y\tilde{g}(\mathcal{Y}(T))$, where $\tilde{g}$ is the dual terminal reward function. The function $\tilde{H}$ is defined as in (\ref{gen_ham}) but with the dual drift and diffusion functions. We use the same notation for the solutions here for brevity, but when considering both primal and dual 2BSDEs simultaneously a distinction will be made. We then discretise and optimise in the same way, but with the addition of a third parameter set $\Theta_3 = \{y\}$ with corresponding loss function 
\begin{comment}
Set $\mathcal{Y}_0 = y^0_0,\, V_0 = V_0^0$ and $Z_0 = Z_0^0$. Then for $i = 0, \ldots N-1$ let $\beta_{i} = N_{\theta_i^0}(\mathcal{Y}_{i})$, $\Gamma_{i} = N_{\lambda_i^0}(\mathcal{Y}_{i})$ and
\begin{align}
\begin{split}
\mathcal{Y}_{i+1} & = \mathcal{Y}_i + (t_{i+1}-t_i)\tilde{b}(t_i, \mathcal{Y}_i,\beta_i) + \sqrt{t_{i+1}-t_i} \tilde{\Sigma}(t_i, \mathcal{X}_i,\beta_i)dW_i \\
V_{i+1} & = V_i + \sqrt{t_{i+1}-t_i} Z_i^\top \tilde{\Sigma}(t_i, \mathcal{Y}_i,\beta_i) dW_i \\
Z_{i+1} & = Z_i - (t_{i+1}-t_i)D_x \tilde{H}(t_i,\mathcal{Y}_i, \beta_i, Z_i, \Gamma_i \tilde{\Sigma}(t_i, \mathcal{Y}_i, \beta_i)) + \sqrt{t_{i+1}-t_i} \Gamma_i \tilde{\Sigma}(t_i, \mathcal{Y}_i,\beta_i) dW_i,
\end{split} \label{2BSDE_disc_dual}
\end{align}
we have analogous loss functions $\LL_1$ and $\LL_2$ for this scheme. Optimising these solves the dual value problem. We now introduce a third loss function to recover the primal value function from this system, as in (\ref{alt_value_def})
\end{comment}
\begin{align}
\LL_3(y) = \E\left[\tilde{g}(\mathcal{Y}_N)\middle|\mathcal{Y}_0 = y\right] + yx_0. \label{loss_y}
\end{align}
%We cannot use the dual value function approximation at time 0, given by $V_0$, since it does not depend explicitly on $y$. We could alternatively find a functional representation (in the style of \cite[Subsection~2.2]{hure2018deep2}) of $V_0$ in terms of $y$, then optimise against this function. The effectiveness of this method is left to further research.

\subsection{Numerical Examples}

In this section we apply the deep controlled 2BSDE algorithm to solve utility maximisation, with varying utilities and control constraints. This is a control problem (\ref{value}) with $d = 1$ and $n = m$. For the primal problem, we have the process $\mathcal{X} = X$ defined by (\ref{wealth}), controlled by $\alpha = \pi$, and the primal 2BSDE system of Theorem \ref{2BSDE_thm} denoted by $(V_1,Z_1,\Gamma_1)$, where we take $g(x) = U(x) \ind_{x>0}$ (c.f. Remark \ref{terminal}) and define the drift and diffusion by
\[ b(t,x,\pi)  = x(r(t) + \pi^\top \left(\mu(t) - r(t)\ind_m\right)), \quad
\Sigma(t,x,\pi) = x \pi^\top \sigma(t).
\]
%\begin{align}
%\begin{split}
%dX(t) & = X(t)  \left(r(t) + \pi(t)^\top(b(t) - r(t)\ind_M)\right) dt + X(t)\pi(t)^\top \sigma(t) dW(t) \\
%dV_1(t) & =  X(t) Z_1(t) \pi(t)^\top \sigma(t) dW(t) \\
%dZ_1(t) & = - \left[ \left(r(t) + \pi(t)^\top (b(t) - r(t) \ind_m )\right)Z_1(t) +  \left| \sigma(t)^\top\pi(t)\right|^2 X(t) \Gamma_1(t) \right] dt \\
%& + \Gamma_1(t) X(t) \pi(t)^\top \sigma(t)dW(t) \\
%X(0) & = x_0 \\
%V_1(T)  & = U(X(T)) \\
%Z_1(T) & = U'(X(T)).
%\end{split} \label{utility_2BSDE}
%\end{align}
For the dual problem the state process $\mathcal{Y} = Y$ is defined in (\ref{Yupdate}), controlled by $\beta = v$ and starting at $y_0 > 0$, and the 2BSDE system of Theorem \ref{2BSDE_thm} is denoted by $(V_2,Z_2,\Gamma_2)$, where  we take $\tilde{g}(y) = \tilde{U}(y)\ind_{y>0}$ and define the drift and diffusion by
\[
\tilde{b}(t,y,v)  = -y(r(t) + \delta_K(v)), \quad
\tilde{\Sigma}(t,y,v) = -y(\theta(t) + \sigma(t)^{-1}v)^\top.
\]
%\begin{align}
%\begin{split}
%dY(t) & = -Y(t)\big(r(t) + \delta_K(\hat{v}(t))\big) dt -Y(t) \left(\theta(t) +\sigma(t)^{-1}\hat{v}(t)\right)^\top dW(t)  \\
%dV_2(t) & = - Z_2(t)Y(t)\left[\theta(t) +\sigma^{-1}(t)\hat{v}(t)\right]^\top dW(t) \\
%dZ_2(t) & =  \left[  \big(r(t) + \delta_K(\hat{v}(t))\big)Z_2(t) + \left|\theta(t) +\sigma^{-1}(t)\hat{v}(t)\right|^2 Y(t) \Gamma_2(t)  \right]dt   \\
%& - \Gamma_2(t) Y(t) \left(\theta(t) +\sigma^{-1}(t)\hat{v}(t)\right)^\top dW(t)  \\
%Y(0) & = y \\
%V_2(T) & = \tilde{U}(Y(T)) \\
%Z_2(T) & = \tilde{U}'(Y(T))
%\end{split} \label{utility_2BSDE_dual}
%\end{align}
We use the following primal-dual relations, derived in \cite{li2018dynamic}.
\begin{theorem}[\cite{li2018dynamic}, Theorem~12] \label{utility_relations_thm} Suppose that $\hat{y}$ and $\hat{v}$ are optimal dual controls, with the corresponding dual state process $\hat{Y}$ and 2BSDE solutions $(\hat{V}_2, \hat{Z}_2,\hat{\Gamma}_2)$. Then the primal state process $\hat{X}$ and 2BSDE solutions $(\hat{V}_1, \hat{Z}_1, \hat{\Gamma}_1)$ corresponding to the optimal primal control $\hat{\pi}$ satisfy
\begin{equation}
\hat{X}(t)  = - \hat{Z}_2(t), \quad
\hat{V}_1(t)  = \hat{V}_2(t) + \hat{Z}_2(t)^\top \hat{Y}(t), \quad
\hat{Z}_1(t)  =  \hat{Y}(t).
\label{utility_relations}
\end{equation}
\end{theorem}
The converse relation also applies, in that primal solutions can be used to derive the corresponding dual solutions. The direction described here is often the most useful, as the dual problem can be  simpler to solve than the primal problem. In addition to these processes, formulas for the optimal controls are also given, but we do not use them here. To show the effectiveness of the algorithm we apply it to a series of utility problems, with certain constraints on the control space.

\begin{example}[Unconstrained Non-HARA Utility Problem] \label{HARA}  We start with the unconstrained case $K = \R^m$, corresponding to a complete market. We consider the utility problem where $U$ has the following non-HARA form
\begin{align}
U(x) = \frac{1}{3} H(x)^{-3} + H(x)^{-1} + xH(x), \qquad x>0, 
\label{nonHARA}
\end{align}
where $H(x) = \sqrt{2}\left(-1 + \sqrt{1+4x}\right) ^{-\half}.$ The  dual utility is given by
$\tilde{U}(y) = \frac{1}{3}y^{-3} + y^{-1}.$
We have $\tilde{K} = \{0\}$, so the optimal dual control is $v(t) = 0$. Suppose that $r(s) = r,\, \sigma(s) = \sigma, \, \mu(s) = \mu$ and $\theta(s) = \theta$ are constant. It is shown in \cite{li2018dynamic}  that the optimiser $\hat{y}$ and processes $\hat{Y}$ and $\hat{Z}_2$ are given by
\begin{align*}
\hat{y} & = \frac{1}{\sqrt{2x_0}}\left[e^{(r+|\theta|^2)T} + \sqrt{e^{2(r+|\theta|^2)T} + 4x_0e^{3(r+2|\theta|^2)T}}\right]^{\half} \\
\hat{Y}(t) & = \hat{y}  \exp\left(-\left( r + \half |\theta|^2  \right)t - \theta^\top dW(t) \right) \\
\hat{Z}_2(t) & = a_1S_1(t) + a_2S_2(t),
\end{align*}
where
$a_1 = \hat{y}^{-4}e^{3(r+2|\theta|^2)T} $, $a_2 = \hat{y}^{-2}e^{(r+|\theta|^2)T}$, $S_1(t) = e^{(r-4|\theta|^2)t + 2\theta^\top W(t)}$ and $S_2(t)  = e^{rt + 2\theta ^\top W(t)}.
$ Furthermore, the dual value function is given by
\[\tilde{u}(t,y) = \frac{1}{3}y^{-3}e^{(3r + 6|\theta|^2)(T-t)} + y^{-1} e^{(r+|\theta|^2)(T-t)}. \]
From these analytical formulae, the primal processes can be derived using Theorem \ref{utility_relations_thm}. For our application, we take $m = 5$, $T = 0.5$, $x_0 = 1$, $r = 0.05$, $\mu = 0.06 \ind_5$ and set $\sigma$ as a random matrix of positive numbers between 0 and 0.2 to introduce correlation of the underlying Brownian motions, with 0.2 added to the diagonal to ensure invertibility.

Figure \ref{utility_whole_loss} shows the development of the loss functions for the primal and dual algorithms with 50 time steps. For the control, we want to maximise the Hamiltonian, which means the derivative of the loss function should go to 0. Hence, we plot the function
$j \mapsto \sum_{i = 0}^{49} \sum_{\theta \in \theta^j_i} |D_\theta \LL (\theta_i^j, i)| $
using the notation of (\ref{L_2}), where $\theta_i^j$ denotes the parameter set for the neural network forming $\pi_i$, at iteration step $j$ of the algorithm. For the dual graph we plot the derivative of $\LL_3$ with respect to $y$, since the above function is always zero as the dual control is forced to be zero, hence optimal, by the support function $\delta_K$.
Over the 10000 iteration steps both losses become small and settled, and therefore are deemed to have converged. In this method we choose an initial learning rate of $10^{-2}$ for the 2BSDE algorithm and the start control $y$ in the dual algorithm, and $10^{-3}$ for the control part. Over the iterations, we divide these rates by 10 three times at regular intervals. We repeat this choice of learning rates for all applications in the sequel, though the number of iteration steps may change.

Figure \ref{utility_whole_processes} shows two simulations of the processes $V_1$ and $Z_1$ (labelled as primal), along with their dual counterparts given by the relations (\ref{utility_relations}) using 20 time steps. We plot these, along with the explicit solution for reference. Both primal and dual approximations are very close to the solutions. The dual state process, appearing in the $Z_1$ graph (but very close to the solution process), is closer than the corresponding primal process since there is no error for the dynamic dual control.

Figure \ref{utility_whole_value} shows the evolution of the approximation of the initial value $u(0,x_0)$ against number of time steps for the primal and dual algorithms, where $x_0 = 1$. The dual value is higher than the primal value, as expected, and the gap between these shrinks as the number of time steps increases. The final relative approximation error is $7 \times 10^{-4}$\% for the primal problem, and $3 \times 10^{-4}$\% for the dual problem, showing high accuracy of the proposed scheme at time 0.
\begin{figure}[h] 
\centering
\begin{minipage}{.5\textwidth}
\centering
\begin{subfigure}[b]{\textwidth}
\includegraphics[width=\textwidth]{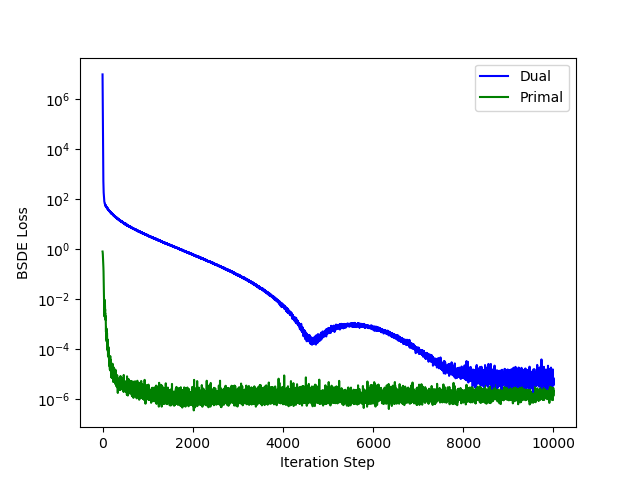}
\caption{Loss function of the BSDE.}
\end{subfigure}
\end{minipage}%
\begin{minipage}{.5\textwidth}
\centering
\centering
\begin{subfigure}[b]{\textwidth}
\includegraphics[width=\textwidth]{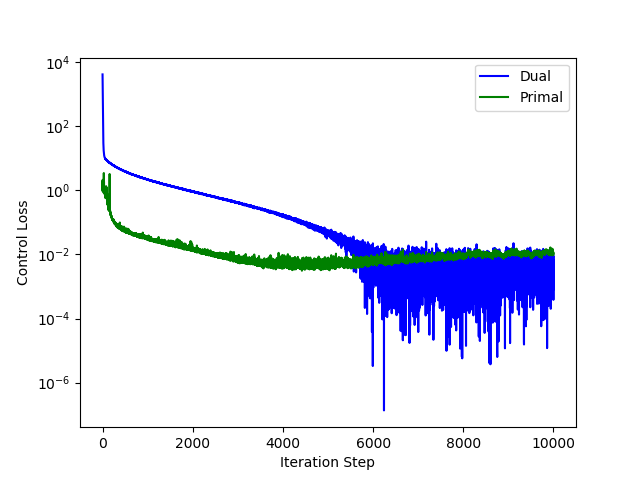}
\caption{Derivative of the control loss function.}
\end{subfigure}
\end{minipage}
\caption{Loss functions against iteration step for the primal and dual deep controlled 2BSDE methods applied to the unconstrained non-HARA utility problem.}

\label{utility_whole_loss}
\end{figure}

\begin{figure}[h]
\centering
\begin{minipage}{.5\textwidth}
\centering
\begin{subfigure}[b]{\textwidth}
\includegraphics[width=\textwidth]{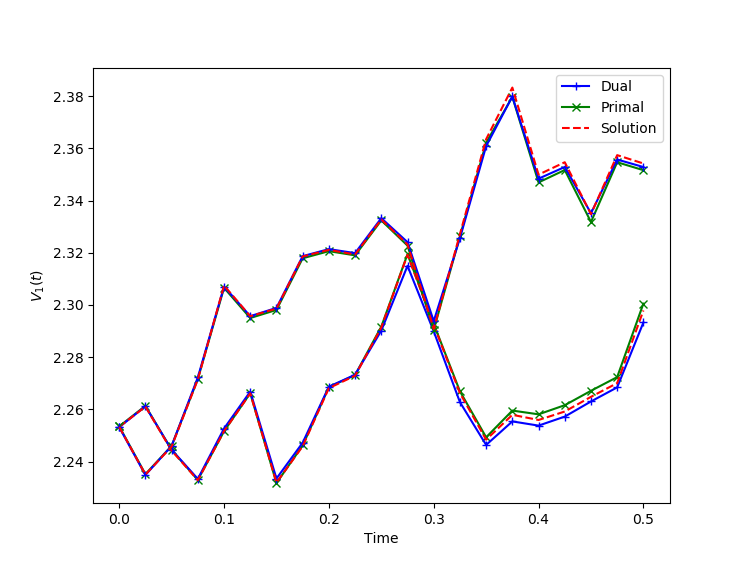}
\caption{Simulation of the process $V_1$.}
\end{subfigure}
\end{minipage}%
\begin{minipage}{.5\textwidth}
\centering
\centering
\begin{subfigure}[b]{\textwidth}
\includegraphics[width=\textwidth]{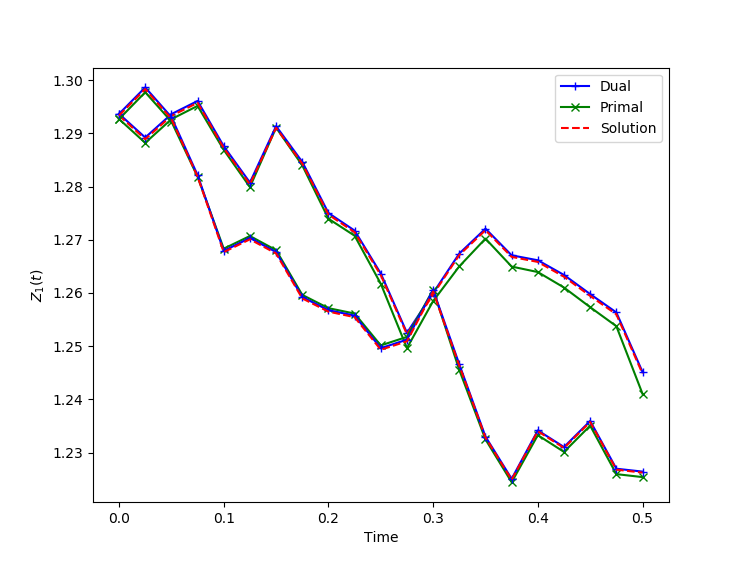}
\caption{Simulation of the process $Z_1$.}
\end{subfigure}
\end{minipage}
\caption{Two simulations of the value process and value-derivative process for the primal and dual deep controlled 2BSDE methods applied to the unconstrained non-HARA utility problem. The displayed dual processes are those implied by the duality relations (\ref{utility_relations}).}
\label{utility_whole_processes}
\end{figure}

\begin{figure}[h]
\centering
\begin{minipage}{.5\textwidth}
\centering
\begin{subfigure}[b]{\textwidth}
\includegraphics[width=\textwidth]{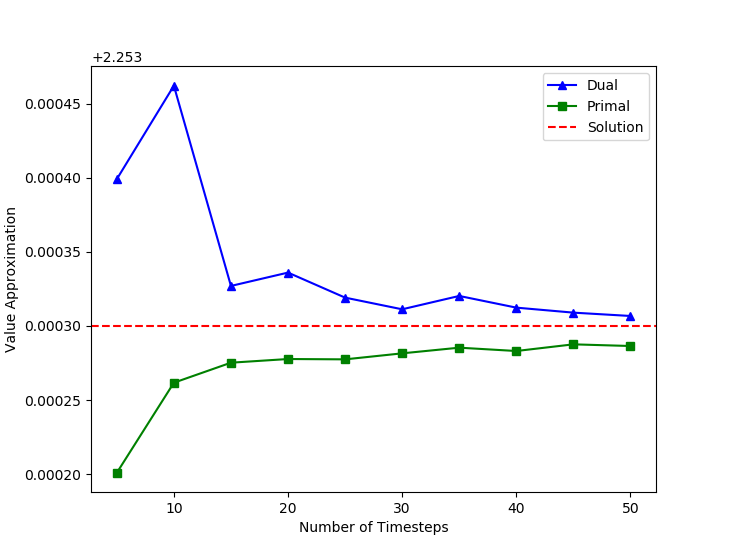}
\caption{Value approximation.}
\end{subfigure}
\end{minipage}%
\begin{minipage}{.5\textwidth}
\centering
\centering
\begin{subfigure}[b]{\textwidth}
\includegraphics[width=\textwidth]{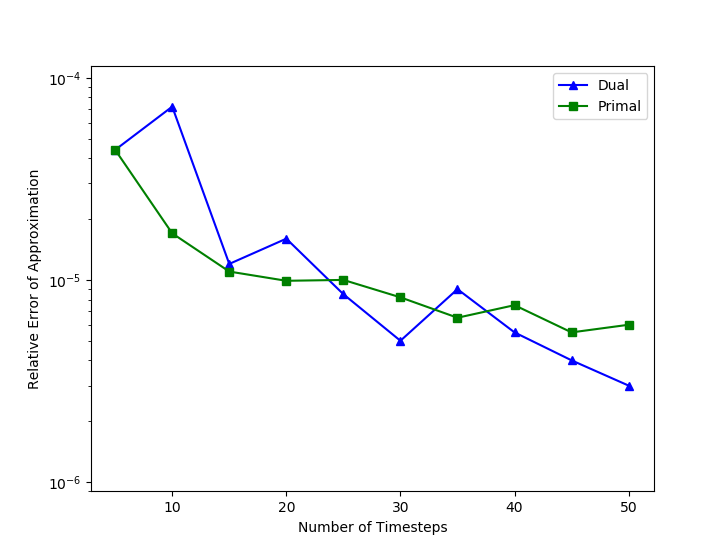}
\caption{Relative error of value approximation.}
\end{subfigure}
\end{minipage}
\caption{Approximation and relative error of the value function against number of time steps $N$ for the primal and dual deep controlled 2BSDE methods applied to the unconstrained non-HARA utility problem. Note in the left graph that the Y values on the axis are shifted down by 2.253.}
\label{utility_whole_value}
\end{figure}

\end{example}

\begin{example}[Cone-Constrained Merton Problem] \label{cone}
Now we consider the problem with power utility $U(x) = \frac{1}{p} x^p$ for some $0<p<1$  when the control space is the positive cone $K = \R_+^m$, corresponding to no short selling constraints.
Let $\hat{X}$ be the primal state process and $(\hat{V}_1,\hat{Z}_1,\hat{\Gamma}_1)$ the solutions of (\ref{2BSDE}) for the primal problem with optimal control $\hat{\pi}$, and let $\hat{Y}$ and $(\hat{V}_2,\hat{Z}_2,\hat{\Gamma}_2)$ be the state process and solutions respectively for the dual problem with optimal controls $\hat{v}$ and $\hat{y}$. The dual state process is a geometric Brownian motion given by
\[\hat{Y}(t) = \hat{Y}(0) \exp\left(-\int_0^t \left( r(s) + \half |\hat\theta(s)|^2  \right) ds - \int_0^t \hat\theta(s)^\top dW(s) \right) \]
in terms of the unknown start point $\hat{y} = \hat{Y}(0)$. The corresponding dual utility function is
$\tilde{U}(y) = \frac{1-p}{p}y^{\frac{p}{p-1}}. $
In \cite[Subsection~4.1]{li2018dynamic} the following solutions are found:
\begin{align}
\begin{split}
\hat{y} & = x_0^{p-1} \exp \left( (1-p) \int_0^T \left[ \frac{p}{2(p-1)^2} |\hat{\theta}(s)|^2 - \frac{p}{p-1}r(s)\right] ds \right) \\
\hat{Z}_2(t) & = -x_0 \exp\left( \int_0^t \left[ r(s) + \frac{1-2p}{2(1-p)^2}|\hat{\theta}(s)|^2 \right] ds + \frac{1}{1-p} \int_0^t \hat{\theta}(s)^\top dW(s) \right), 
\end{split} \label{merton1}
\end{align}
where $\hat{\theta}(t) = \theta(t) + \sigma(t)^{-1}\hat{v}(t)$. This deterministic optimal dual control can be found numerically using Python packages such as \texttt{Scipy}.
The dual value function is given by
\begin{align}
\tilde{u}(t,y) = \tilde{U}(y) \exp\left(\int_t^T \left[ \frac{p}{2(p-1)^2}|\hat{\theta}(s)|^2  - \frac{p}{p - 1} r(s) \right] ds \right). \label{merton2}
\end{align}

For our implementation, we take $T = 0.5$, $x_0 = 1.0$, $m = 50$, $r(t) = 0.05$ and $p = \half$. We define the drift as $\mu(t)_i = 0.04 + \sin(\pi t + A_i) / 50$ where $A_i$ is randomly generated for each $i=1,\ldots,50$. We choose this function so that in the unconstrained case it would be optimal to short sell at some points. We set $\sigma(t)$ to be a constant matrix with 0.4 on the diagonal, and 0.2 everywhere else.

\begin{remark}
To ensure our controls $\pi$ and $v$ are in the correct set $K = \tilde{K}  =\R_+^m$, we (element-wise) apply the function $x \mapsto \max(0,x)$ to the output of their respective neural networks. This function is an almost everywhere differentiable surjection from $\R^m$ to $\R_+^m$, which is sufficient for using a gradient descent method to find the optimal control. The mapping is not injective, but this is not an issue as the exact output of the neural network is not relevant. We care only about the control outputted by projecting onto the control space.
\end{remark}
\end{example}

\begin{figure}[h]
\centering
\begin{minipage}{.5\textwidth}
\centering
\begin{subfigure}[b]{\textwidth}
\includegraphics[width=\textwidth]{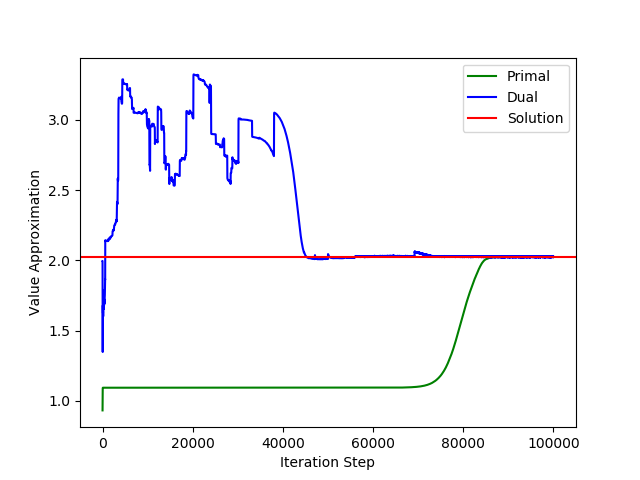}
\caption{Value approximation.}
\end{subfigure}
\end{minipage}%
\begin{minipage}{.5\textwidth}
\centering
\centering
\begin{subfigure}[b]{\textwidth}
\includegraphics[width=\textwidth]{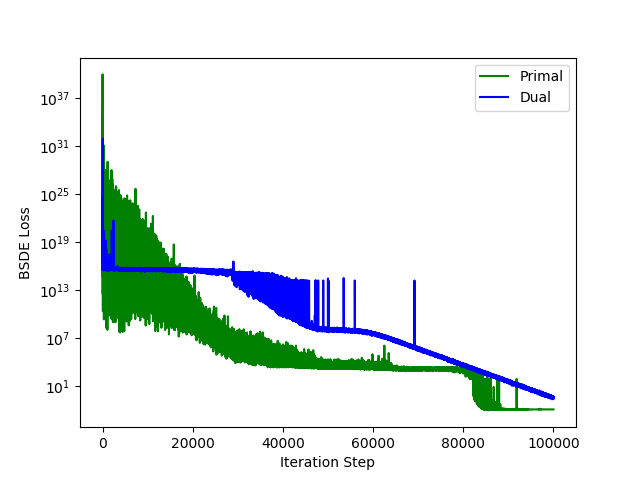}
\caption{BSDE loss.}
\end{subfigure}
\end{minipage}
\caption{Approximation of the value function and bsde loss against iteration step for the primal and dual deep controlled 2BSDE methods applied to the cone constrained power utility problem.}
\label{pow_cone}
\end{figure}

Figure \ref{pow_cone} shows the results of applying the primal and dual 2BSDE methods to this problem. We use $N = 10$ time steps and run the algorithm for 100000 steps, notably more than for the lower dimensional unconstrained problem. This increases the run time to 5457 seconds. We start the BSDE and control learning rates at 0.01 and 0.001 respectively, then reduce these by a factor of 10 half way through training. We run both algorithms at the same time, which saves time as we can use the same generated Brownian motions for each algorithm. We see less convergent structure at the beginning of training, perhaps due to bad initial guesses for the neural networks. Indeed, we see that the value for the primal functions stays nearly constant until the loss function drops to a certain level, when the processes are accurate enough that the value approximation can move towards the analytical solution. The dual algorithm behaves similarly, but with more noise. At the end of training we have relative errors of 0.1\% and 0.05\% for the primal and dual algorithms respectively. This error is a few orders of magnitude higher than the unconstrained case due to the increase in dimension.

\begin{example}[Ball Constrained Log Utility Problem] \label{log} In this example we consider the log utility problem when the control process is constrained to the ball $B(0,R)$ of radius $R > 0$ in $\R^m$. This corresponds to the investor not wanting to invest more than a certain proportion of their wealth in risky stocks. In this case, the support function $\delta_K(z) = R|z|$ does not vanish for all admissible dual controls. Therefore, the SDE (\ref{Yupdate}) is harder to solve. In general this problem is also difficult to solve analytically because of the dependence of the dual controls $\hat{y}$ and $ \hat{v}$ on each other. However, we can exploit the multiplicity property of the log function to decouple these controls to give analytical solutions, which we describe here. To this end let $U(x) = \log(x)$ be our utility function, for which the dual utility is $\tilde{U}(y) = -(1 + \log(y))$. Since the dual process $Y$ is a geometric Brownian motion we have
\[ \E\left[\tilde{U}(Y(T))\right] = -1 - \log(Y(0)) + \int_0^T r(t) + R|v(t)| + \half |\theta(t) + \sigma(t)^{-1} v(t)|^2 dt. \]
The measurable process
\[\hat{v}(t) \defeq \arg \min_{v \in \R^m} \left\{ R|v| + \half |\theta(t) + \sigma(t)^{-1} v|^2 \right\} \]
is an optimal control, and can be found using existing deterministic convex optimisation algorithms. Independently of this, the dual control $\hat{y}$ is found via
\[\hat{y} \in \arg\min_{y >0} \left\{ x_0y - \log(y) + \int_0^T r(t) + \delta_K(\hat v(t)) + \half |\theta(t) + \sigma(t)^{-1} \hat v(t)|^2 dt\right\}. \] 
yielding $\hat{y} = \frac{1}{x_0}$. The value function is then given by
\begin{align}
u(0,x_0) =  \log(x_0) + \int_0^T r(t) + R|\hat{v}(t)| + \half |\theta(t) + \sigma(t)^{-1} \hat{v}(t)|^2 dt. \label{log_sol}
\end{align}
For our application, we  take $R = 1.0$, $T = 0.5$, $x_0 = 5.0$, $m = 20$, $r(t) = 0.05$ and $\mu(t) = 0.07\ind_5$. We define the volatility as a diagonal matrix with the $i$th diagonal element equal to
$\sigma(t)_i = 0.4 + 0.2\sin(2\pi t + A_i)$
for some arbitrarily (in our case randomly) chosen numbers $A_i$ to shift the peaks of the sine curve for each dimension. We choose this function so that in the unconstrained case it would be reasonable to invest more than we are allowed at points when the volatility is low. The volatility is non-negative for all $t \in [0,T]$, so the problem is well posed. To ensure that our primal control is in the correct set, we use a penalty function method. In this case we do not constrain $\pi$, but instead we alter the loss function (\ref{L_2}) to be
\[ \LL_2(\theta_i,i) \defeq \E\left[ F( t_i, X_i,\pi_i, Z_i, \Gamma_i) - \beta \max(0,|\pi_i| - 1)^2\right] \]
which  penalises the control for straying outside of $K$. The parameter $\beta > 0$ is some large constant, set to 1000 for our example. 
%We find that this method produces a more accurate value approximation than using hard constraints. 
%For example, using a function that rescales the norm of $\pi$ via the mapping $x \mapsto e^{ - |x|^2}$, we achieve a relative error of 0.8\% with the same runtime, as opposed to 0.7\% using soft constraints. We do not use such a function for the dual control, as it is never infinitely penalised for its position. 

Table \ref{log_tab} shows the evolution of the approximated value function for the primal and dual algorithms, compared with the numerical solution 1.64755570 found using (\ref{log_sol}). The algorithm is run using $N = 10$ time steps, and the value approximation, relative error and (combined) runtime are recorded. We again perform both algorithm simultaneously using the same Brownian motions simulations. The value approximation initially moves away from the solution, but then converges.  For the primal algorithm, the value moves away over the first 1000 steps, then goes to the solution. For the dual algorithm, the value does not move as far from the solution, but takes longer to do so, only starting moving towards the solution at 10000 steps. The dual value initially dips below the solution, then increases, which is why the relative error is lower at 1000 steps than at the start. 

\begin{table}[h]
\centering
\begin{tabular}{c||c|c||c|c||c}
Step & Primal Value & Rel-Err (\%) & Dual Value & Rel-Err (\%) & Time (s) \\
 \hline  \hline
1 & 9.92774e-01 & 3.97426e+01  & 2.50412e+00 & 5.19898e+01 & 128 \\
1000 & 4.63730e-02 & 9.71853e+01  & 1.57275e+00 &  4.54015e+00 & 183 \\
5000 & 2.27800e-01 & 8.61735e+01 & 1.80325e+00 & 9.44972e+00 & 405 \\
10000 & 1.39359e+00 & 1.54147e+01 & 2.01337e+00 & 2.22035e+01 & 701 \\
12000 & 1.62170e+00 & 1.56918e+00 & 1.98620e+00 & 2.05544e+01 & 815 \\
18000 & 1.63545e+00 & 7.34658e-01 & 1.68103e+00 & 2.03169e+00 & 1149 \\
%19000 & 1.63546e+00 & 7.34163e-01 & 1.65375e+00 & 3.76032e-01 & 1204 \\
20000 & 1.63543e+00 & 7.35726e-01 & 1.65361e+00 & 3.67465e-01 & 1259
\end{tabular}
\caption{Value approximation and runtime at certain iteration steps for the primal and dual deep controlled 2BSDE methods applied to the ball-constrained log utility problem.}
\label{log_tab}
\end{table}
\end{example}

\section{The Deep SMP Method for the Non-Markovian Case} \label{Non-Markovian}

In this section we consider the case in which  $r$, $\mu$ and  $\sigma$ may not be deterministic. The utility  problem and its dual problem then become non-Markovian. This means we cannot write the value function as a measurable function of state and time nor apply the dynamical programming principle, hence the HJB equation is invalid. However, we can still employ the SMP, and it is shown in \cite{li2018dynamic} that there exist certain optimality conditions for a non-Markovian utility maximisation problem, which we use to define a new algorithm, named the deep SMP method. In some of the examples we consider in this section, the problem can be made Markovian by introducing some new artificial structure to the problem. In these cases, we  compare the accuracy and runtime of both algorithms.

\subsection{The Deep SMP Method}

Consider an agent who invests in the market (\ref{market}) in full generality. We again consider the utility problem, written in this setting as
\[u(t) = \sup_{\pi \in {\cal A}} \E \big[ U(X(T)) \, \big| \, \F_t  \big]. \]
We cannot use It\^o's formula for this value process as it cannot be written as a measurable function of the wealth process $X(t)$, defined as in (\ref{wealth}) and controlled by $\pi$, but we can still use the SMP. For the primal problem we have the following adjoint BSDE with unknown processes $P_1$ and $Q_1$ valued in $\R$ and $\R^m$ respectively:
\begin{equation}
dP_1(t)  = - \left[ \left( r(t) + \pi(t)^\top \left(\mu(t) - r(t)\ind_m\right) \right)P_1(t) + Q_1(t)^\top\sigma(t)^\top\pi(t)\right]dt + Q_1(t)^\top dW(t)  \label{utility_BSDE}
\end{equation}
with the terminal condition $P_1(T)  = - U'(X(T))$. Similarly, we can define the dual state $Y$ as per (\ref{Yupdate}), controlled by $(y,v)$. The dual adjoint equation is the following BSDE with unknown processes $P_2$ and $Q_2$ valued in $\R, \R^m$ respectively:
\begin{equation}
dP_2(t)  = \left[(r(t) + \delta_K(v(t)) P_2(t) + Q_2(t)^\top (  \theta(t) + \sigma(t)^{-1}v(t) 
) \right] dt + Q_2(t)^\top dW(t) \label{utility_BSDE_dual}
\end{equation}
with the terminal condition $P_2(T)  = -\tilde{U}'(Y(T))$.  Furthermore, we have the duality relations $P_2(t) = X(t)$ and $P_1(t) = -Y(t)$ almost surely, for almost every $t \in [0,T]$ by (\ref{utility_relations}). We now state the optimality conditions for the dual problem in terms of these processes.
\begin{theorem}[\cite{li2018dynamic}, Theorem 11] \label{dual_opt} Let $(y,v)$ be an admissible dual control. Then $(y,v)$ is optimal if and only if the solutions ($P_2,Q_2$) of (\ref{utility_BSDE_dual}) satisfy the following
\begin{enumerate}
\item $P_2(0) = x_0$,
\item $\frac{(\sigma(t)^\top)^{-1}Q_2(t)}{P_2(t)} \in K$,
\item $P_2(t)\delta_K(v(t)) + Q_2(t)^\top \sigma(t)^{-1}v(t) = 0 $
\end{enumerate}
for all $t \in [0,T]$ almost surely.
\end{theorem}
\begin{remark} \label{primal_opt} The corresponding optimality condition for a primal control $\pi$ with corresponding state process $X$ and adjoint BSDE solutions $(P_1, Q_1)$ is
\[ [\pi(t) - a]^\top [X(t)\sigma(t)(P_1(t)\theta(t) + Q_1(t))] \geq 0 \]
for all $t \in [0,T]$ and $a \in K$ almost surely. This is a much more complex condition, and the implementation would require some grid-based method to consider a sufficiently large number of $a \in K$, which would be computationally costly, especially in higher dimensions and unbounded (but still constrained) $K$. For this reason, we approach the dual conditions and then use the primal dual relations to derive the necessary primal processes.
\end{remark}
The machine learning methodology to solve this problem is as follows. First, we simulate the dual processes in the forward direction using the Euler-Maruyama scheme. We start the process $P_2$ at $P_2(0) = x_0$ to ensure the first condition is satisfied. This condition is why it appears that the forward simulation method is naturally suited to this problem. We use neural networks to define the processes $v$ and $Q_2$, then update the values of $Y$ and $P_2$ accordingly. Specifically, we set $Y(0) = y$, $P_2(0) = x_0$ then for $i = 0, \ldots , N-1$ set  
$v(i)  = N_{ \theta^v_i}\left(Y(i)\right)$ and
\begin{align}
\begin{split}
Q_2(i) & = P_2(i) \sigma(t_i)^\top h_K\left(N_{\theta^Q_i}\left(Y(i)\right)\right)\\
Y(i+1) & = Y(i) -(t_{i+1} - t_i)Y(i)\big(r(t_i) + \delta_K(v(i))\big)  - Y(i) \left(\theta(t_i) + \sigma(t_i)^{-1}v(i)\right)^\top dW_i \\
P_2(i+1) & = P_2(i) + (t_{i+1} - t_i)\left[r(t_i)P_2(i) + Q_2(i)^\top  \theta(t_i)  \right]  + Q_2(i)^\top dW_i, \label{SMP_disc}
\end{split}
\end{align}
where $\theta^v_i, \theta^Q_i \in \R^{\rho(d,m)}$ are parameters for neural networks, $dW_i$ is a multivariate normal $\Norm_m(0,(t_{i+1} - t_i) \ind_m)$ random variable, and $h_K \colon \R^m \to K$ is a differentiable surjective function that ensures condition 2 in Theorem \ref{dual_opt} is  satisfied. The two neural networks used here both output to $m$-dimensional space, unlike the general algorithm outlined in Section \ref{ML}, where the second neural network is valued in $\R^{d \times d}$. 
%This increases the number of parameters, as for the utility maximisation problem we have $d = 1$. However, as we move into the case of random coefficients that satisfy their own SDEs, this dimensionality also increases. 
We have the variables $y, \Theta^Q \defeq (\theta^Q_i)_{i=0}^{N-1}$ and $\Theta^v \defeq (\theta_i^v)_{i=0}^{N-1}$, which are optimised against the following loss functions:
\begin{align*}
\LL^y(y) & =  \E\left[\tilde{U}(Y(N))\middle|Y(0) = y\right] + x_0 y \\
\LL^Q(\Theta^Q) & = \E\left[|\tilde{U}'(Y(N)) + P_2(N)|^2\right] \\
\LL^v(\theta^v_i) & = \E\left[|P_2(i) \delta_K(v(i)) + Q_2(i)^\top\sigma(t_i)^{-1} v(i) |^2\right].
\end{align*}
We take sample means to approximate the expected values.  The first loss function represents the optimality condition of $y$ from the definition (\ref{duality}) of the dual problem, the second  the terminal condition of BSDE (\ref{utility_BSDE_dual}),   and  the third condition~3 of Theorem \ref{dual_opt}. 

The optimisation process is very similar to the deep controlled 2BSDE method as outlined in Section \ref{ML}, so we only provide a brief summary here. For each iteration step we simulate the processes $Y$, $v$, $P_2$ and $Q_2$ using our existing (or randomly initialised for the first step) parameter sets $y$, $\Theta^Q$ and $\theta^v$ and a batch of simulated Brownian motion paths. We then perform one iteration of the \texttt{ADAM} algorithm to update each parameter set in turn using their respective loss functions. In between the updates, we simulate the processes again using the new parameters, but do not generate new Brownian motions. When the 3 sets are updated we generate new Brownian motion paths and repeat. This process continues until the difference made by updating is sufficiently small. For our case, we start with a learning rate of 0.01 for all parameters, which decreases as the number of iterations increases.

In this algorithm we no longer have a variable representing the value at time 0. We therefore need to perform Monte Carlo simulations, forming bounds of the value function. An upper bound can be found using the dual state process $Y$ and a lower bound the primal state process $X = P_2$, as per (\ref{utility_relations}). The estimates outputted are given by
\begin{align}
u_\text{low} \defeq \frac{1}{M} \sum_{i=1}^M U(P_2^i(N)) \leq u(0,x_0) \leq \frac{1}{M} \sum_{i=1}^M \tilde{U}(Y^i(N)) + x_0y \eqdef u_\text{high} \label{SMP_val}
\end{align}
for some large batch of size $M \in \N$, where each path is indexed by the superscript $i$. The start point $y$ is not random as it is optimised using a deterministic loss function. The inequalities are valid up to discretisation and simulation errors, and as the number of time steps increases the gap $u_\text{high} - u_\text{low}$ decreases, and the optimal value is eventually contained in this gap.

\subsection{Numerical Examples}

\begin{example}[Power Utility in a Stochastic Volatility Model]

In this example we consider an extension of our utility model to stochastic volatility. In this model, our agent invests in a 2-dimensional stock market, out of which one asset $S$ is traded, and is defined by
\[dS(t) = S(t)(r + Av(t))dt + S(t) \sqrt{v(t)} dW^s(t). \]
The constants $r, A>0$ represent the risk-free rate and market price of risk respectively. The volatility process $v$ has the dynamics
\[dv(t) = \kappa (\theta - v(t)) dt + \xi \sqrt{v(t)} dW^v(t). \]
The constant $\kappa > 0$ represents the speed at which the process reverts to the long-time average volatility, given by $\theta > 0$. The constant $\xi > 0$ is a variance parameter. The Brownian motions $W^s$ and $W^v$ are correlated with correlation parameter $\rho  \in [-1 ,1]$. When the agent invests a proportion $\pi$ of their wealth $X$ in this market, their wealth process (\ref{wealth}) evolves as
\[dX(t) = X(t)(r+\pi(t)Av(t))dt + X(t)\pi(t) \sqrt{v(t)}dW^s(t). \]

We  first consider this as a 2-dimensional Markovian problem, then a random coefficient problem. In the Markovian problem the 2-dimensional process $\mathcal{X} = (X,v)$ is the state process controlled by the $K \subset \R$ valued process $\pi$. Consider the maximum utility problem given by
\[u(t,x,v) = \sup_{\pi \in {\cal A}} \E \left[ U(X(T)) \,\middle|\, (X(0),v(0)) = (x,v) \right] \]
for some utility function $U$. This setup fits in the framework of (\ref{value}) with $d = 2$, $n = 2$, $m = 1$ and $g(x,v) \defeq U(x)\ind_{x>0}$ for $(x,v) \in \R^2$. The drift and diffusion functions are given by
\begin{align*}
b(t,x,v,\pi) & = \left( x(r + \pi A v), \kappa(\theta - v) \right)^\top \\
\Sigma(t,x,v,\pi) & = 
\begin{pmatrix}
x \pi \sqrt{v} & 0 \\
\rho \xi \sqrt{v}  & \sqrt{1 - \rho^2} \xi \sqrt{v} 
\end{pmatrix}.
\end{align*}
It is a usual assumption that the drift and diffusion functions are Lipschitz functions of the state and control variable, but this is not the case here due to the square root appearing in the diffusion. This may invalidate certain convergence patterns that we have seen for the utility maximisation problem thus far.
%The value processes of (\ref{2BSDE}) are given by
%\begin{align*}
%dV_1(t) & = Z_1(t)^\top \sigma(t,X(t),v(t),\pi(t))dW(t) \\
%dZ_1(t) & = -D_{(x,v)}H\big(t,X(t),v(t),\pi(t),Z_1(t),\Gamma_1(t)\sigma(t,X(t),v(t),\pi(t))\big)dt + \Gamma_1(t) \sigma(t,X(t),v(t),\pi(t))dW(t) 
%\end{align*}
%where the state derivative of the generalised Hamiltonian $H$ is given by
%\[D_{(x,v)}H(t,x,v,\pi,z,q) = 
%\begin{pmatrix}
%(r+\pi A v)  z_1 + \pi \sqrt{v} (q_{00} + q_{01}) \\
%x A \pi z_1 - \kappa z_2 + \frac{1}{2\sqrt{v}} (x \pi q_{00} + \xi (\rho + \sqrt{1 - \rho^2})(q_{10} + q_{11})) 
%\end{pmatrix}.
%\]
%Here we use $z_i$ and $q_{ij}$ to indicate the elements of $z$ and $q$ respectively. The matrix $q$ here is actually
%\[\Gamma_1(t) \sigma(t,X(t),v(t),\pi(t)) = 
%\begin{pmatrix}
%\Gamma_{00} x \pi \sqrt{v} + \Gamma_{01} \rho \xi \sqrt{v} & \Gamma_{01} \sqrt{1-\rho^2} \xi \sqrt{v}  \\
%\Gamma_{10} x \pi \sqrt{v} + \Gamma_{11} \rho \xi \sqrt{v} & \Gamma_{11} \sqrt{1-\rho^2} \xi \sqrt{v}  
%\end{pmatrix}
%\]
%where $\Gamma_{ij}$ denotes the elements of $\Gamma_1(t)$. 
The dual state process consists of the volatility process $v$ and an $\R-$valued process $Y$ that is controlled by some $\R^2$-valued process $(\eta, \gamma)$, given by
\begin{align}
dY(t)  = -Y(t) \left[\bigg(r + \delta_K\big(\eta(t)\big)\bigg) dt+\left(\frac{\eta(t)}{\sqrt{v(t)}} + \rho \gamma(t) + A \sqrt{v(t)}\right)dW^s(t) - \gamma(t) dW^v(t) \right]. \label{hest_dual}
\end{align}
As before, the dual process starts at some undetermined $y>0$, which forms the dual control together with $\eta$ and $ \gamma$, and is chosen such that the process $XY$ is a super martingale. Now consider the dual control problem
\begin{align}
\tilde{u}(t,y,v) = \inf_{\gamma,\eta} \E\left[ \tilde{U}(Y(T)) \middle| (Y(t), v(t)) = (y,v) \right]. \label{heston_dual}
\end{align}
%The dual terminal function remains the same as the Legendre transform is not affected by the unused additional variable $v(T)$. Indeed, for all $y, \alpha \geq 0$ we have
%\[\tilde{g}(y,\alpha) = \sup_{x,v \geq 0} \left\{ g(x,v) - xy - v\alpha \right\} = \sup_{x \geq 0} \left\{ U(x) - x y \right\} =  \tilde{U}(y).  \]
%The dual HJB equation is given by 
%\begin{align*} \frac{\partial \tilde{u}}{\partial t} = - \inf_{\gamma,\eta} & \left\{-\left(r + \delta_K(\eta)\right)y\frac{\partial \tilde{u}}{\partial y} + \kappa(\theta - v)\frac{\partial \tilde{u}}{\partial v}
%+ \half y^2\left[\left(A\sqrt{v} + \frac{\eta}{\sqrt{v}}\right)^2 + \gamma^2\left(1-\rho^2\right)\right] \frac{\partial^2 \tilde{u}}{\partial y^2} \right. \\
%& \left. + \half \xi^2 v \frac{\partial^2 \tilde{u}}{\partial v^2} + y\left[\gamma \xi \sqrt{v} \left(1 - \rho^2\right) - \left(A v + \eta\right) \xi \rho\right]\frac{\partial^2 \tilde{u}}{\partial y \partial v} \right\}.
%\end{align*}
In our numerical example we consider the unconstrained case $K = \R$. As before, the presence of the $\delta_K$ term means that we must have $\eta$ = 0. This puts the problem in the same form as \cite{ma2020dual}, where only one process $\gamma$ needs to be found.

This setup fits in our framework for a dual problem with $d = 2$, $m = 1$, and $n =2$. This is not a multidimensional duality as the process $v$ appears in both state dynamics, but a single dimensional duality between $X$ and $Y$. The drift and diffusion functions are given by
\begin{align*}
\tilde{b}(t,y,v,\gamma) & = (-yr, \kappa(\theta - v))^\top \\
\tilde{\sigma}(t,y,v,\gamma) & = 
\begin{pmatrix}
-y A \sqrt{v} & \sqrt{1 - \rho^2} y \gamma \\
\rho \xi \sqrt{v} & \sqrt{1 - \rho^2} \xi \sqrt{v} 
\end{pmatrix}.
\end{align*}
When we use the power utility $U(x) = \frac{x^p}{p}$ for  $0<p<1$, exact solutions can be obtained for this problem. A solution is derived in \cite{ma2020dual} by assuming that the value function $u$ has the form
\begin{align}
u(t,x,v) = \frac{x^p}{p} \exp(C(t) + D(t)v)  \label{hest_value}
\end{align}
in terms of two unknown functions $C$ and $D$ that satisfy a system of Riccati type of ordinary differential equations for which numerical solutions can be found. We use $u(t,x,v)$ as a benchmark. For our application we take $r = 0.05$, $\rho = -0.5$, $p = 0.5$, $\kappa = 1$, $\theta = 0.05$, $\xi = 0.5$, $A = 0.5$, $x_0 = 1$, and $v_0 = 0.5$. 

Table \ref{heston} shows the relative error and run time of our algorithms when we use a range of terminal times $T$ and numbers of time steps $N$. We contrast this to the results of \cite{ma2020dual}, in which, with $T = 1.0$ and $N = 100$, upper and lower bounds for the value were found with relative errors of 0.003\% in 2240 seconds and 27600 seconds respectively, albeit for slightly different parameter choices. The upper bound was found faster using a convenient guess of the form the dual control $\gamma$. We are not using 100 time steps for our algorithms due to computer memory constraints. In fact, the 50 time step algorithm is also slower than expected due to a memory bottleneck. For $T = 0.2$ the solution is 2.03289, for $T = 0.5$ it is 2.07559 and when $T=1.0$ the solution is 2.13420. The run time for the $T = 0.5$ and $T = 1.0$ algorithms is higher than for $T = 0.2$ since twice the iteration steps were taken, 10000 compared to 5000 for $T = 0.2$. While the errors for the primal algorithm reduce as $N$ increases, the value does not lie within the duality gap for some terminal times. This may be down to a discretisation error, and would be fixed by taking more time steps. We see convergence of the approximations to the true value as $N$ increases and this convergence seems to be linear, as we saw in Example \ref{HARA}. This convergence is explored further in Section \ref{method}.

\begin{table}[h]
\centering
\begin{tabular}{c|c||c|c|c||c|c|c}
T & N & Primal Value & Rel-Err (\%) & Time (s) & Dual Value & Rel-Err (\%) & Time (s) \\
 \hline  \hline
0.2 & 5 & 2.03310 & 1.01654E-02 & 83 & 2.03378 & 4.35850E-02 & 98 \\
0.2 & 10 & 2.03301 & 5.74161E-03 & 206 & 2.03344 & 2.66047E-02 & 239 \\
0.2 & 20 & 2.03296 & 3.27670E-03 & 591 & 2.03314 & 1.23181E-02 & 604 \\
0.2 & 50 & 2.03289 & 2.81206E-04 & 2849 & 2.03295 & 2.89678E-03 & 2622 \\
\hline
0.5 & 5 & 2.07616 & 2.75979E-02 & 152 & 2.07969 & 1.97584E-01 & 183 \\
0.5 & 10 & 2.07582 & 1.14780E-02 & 362 & 2.07747 & 9.08529E-02 & 411 \\
0.5 & 20 & 2.07572 & 6.36082E-03 & 927 & 2.07670 & 5.38088E-02 & 1030 \\
0.5 & 50 & 2.07562 & 1.55757E-03 & 5070 & 2.07600 & 1.98421E-02 & 4181 \\
\hline
1.0 & 5 & 2.13328 & 4.32410E-02 & 154 & 2.14847 & 6.68466E-01 & 170 \\
1.0 & 10 & 2.13377 & 1.99985E-02 & 362 & 2.14264 & 3.95457E-01 & 400 \\
1.0 & 20 & 2.13399 & 9.92254E-03 & 918 & 2.13715 & 1.38171E-01 & 981 \\
1.0 & 50 & 2.13403 & 7.97535E-03 & 4230 & 2.13548 & 6.01058E-02 & 4395
\end{tabular}
\caption{Accuracy and runtime for the deep controlled 2BSDE method for various terminal times $T$ and numbers of time steps $N$ applied to the unconstrained power utility problem in a stochastic volatility model.}
\label{heston}
\end{table}
Now, let us treat $v$ as some random process, meaning that we can only sample $v$, and not the underlying Brownian motion $W^v$. In this setting we have that both $\theta(t) = A \sqrt{v(t)}$ and $\sigma(t) = \sqrt{v(t)}$ are random processes for the stock market, and we still wish to apply Theorem \ref{dual_opt}. We cannot directly apply this theorem as $v$ is not measurable with respect to the filtration generated by $W^s$. To circumvent this problem, we adopt a market completion method (c.f. \cite{karatzas1998methods} for example) by first constructing an artificial stock driven by $W^v$, adding it to our market, and then making it unavailable for trading. To be specific, we define a new asset price $S^v$ by
\[dS^v(t) = S^v(t)\left(r dt + dW^v(t)\right),  \]
which ensures that $v$ is adapted to the natural filtration of the Brownian motions driving the market. For simplicity we assume that $m = 1$, but this analysis extends naturally to multiple dimensions. We invest in this market with a 2-dimensional portfolio, restricted to the set $K \times \{0\}$, ensuring that our wealth process $X$ is still defined as in (\ref{wealth}), controlled by a 1-dimensional $K$-valued process $\pi$. Now the dual state process (\ref{hest_dual}) is controlled by $y$ and a 2-dimensional dynamic control $(\eta,\gamma)$ valued in $\tilde{K} \times \R$. We would like to show that we can take $\gamma = 0$, as this would return the dual problem to the 1-dimensional control case independent of $W^v$. We want this as we cannot `see' $W^v$, and can only sample $v$ directly at any time. To this end, let $(P_2, Q_2)$ be the dual adjoint solutions. Fix a time $t \in [0,T]$ and let $Q_2(t) = (q_1, q_2)$. By Theorem \ref{dual_opt} (2) we have $P_2(t)^{-1}q_2 = 0$, hence $q_2 = 0$. However, this means that condition (3) becomes
\[P_2(t) \delta_K(\eta(t)) + q_1^\top \sigma(t)^{-1} \eta(t) = 0 \]
which is independent of $\gamma(t)$, leading to a free control. We therefore can take $\gamma = 0$, removing the dependence of the dual state process on $W^v$. Note that this choice of control may not be optimal, and the decision to remove dependence on $W^v$ may lead to a non-zero duality gap.
%Let $(P_1,Q_1)$ be the primal adjoint solutions. By the relations (\ref{utility_relations}), we have $Y(t) = -P_1(t)$ almost surely for almost every $t \in [0,T]$. Writing $Q_1(t) = (Q_{11}(t), Q_{12}(t))$, we have $\gamma(t) = Q_{12}(t)$. The primal optimality condition (\ref{primal_opt}) tells us that the primal control $\pi$ is optimal if and only if for almost every $a \in K$ we have
%\[ 
%\begin{pmatrix}
%\pi(t) - a \\
%0
%\end{pmatrix}^\top \left(X(t)
%\begin{pmatrix}
%\sigma(t) & 0\\
%0 & 1 
%\end{pmatrix}
%\left[
%P_1(t)
%\begin{pmatrix}
%\theta(t) \\
%0 
%\end{pmatrix}
%+ 
%\begin{pmatrix}
%Q_{11}(t) \\
%Q_{12}(t)
%\end{pmatrix} 
%\right]
%\right) \geq 0, \]
%which is independent of $Q_{12}(t)$. We can take $Q_{12}(t) = \gamma(t) = 0$, returning us to our previous 1-dimensional setting.

Table \ref{hest_SMP} shows the accuracy and runtime of the deep SMP method with a range of terminal times and step sizes, when we run it on the same problem formulation as above. Here, $K = \R$, so we take the constraining function $h_K$ to be the identity.  We note that the upper bounds are an order of magnitude less accurate the lower bounds for this problem. This indicates that the implied primal process approximations are more accurate than the dual approximations in this algorithm. Comparing these results to Table \ref{heston}, the SMP method gives roughly the same accuracy of bounds as the dual deep controlled 2BSDE method, but with a lower computation time. In particular, the algorithm here only needed 5000 iterations steps, compared to 10000 for both the 2BSDE algorithms.
\begin{table}[h]
\centering
\begin{tabular}{c|c||c|c||c|c|c}
T & N & $u_\text{low}$ & Rel-Diff (\%) & $u_\text{high}$ & Rel-Diff (\%) & Time (s) \\
 \hline  \hline
0.2 & 5 & 2.03271 & 9.17595E-03 & 2.03348 & 2.90360E-02 & 103 \\
0.2 & 10 & 2.03280 & 4.40981E-03 & 2.03323 & 1.64607E-02 & 273 \\
0.2 & 20 & 2.03284 & 2.60193E-03 & 2.03305 & 7.49044E-03 & 781 \\
0.2 & 50 & 2.03289 & 5.76610E-05 & 2.03295 & 2.54626E-03 & 4788 \\
\hline
0.5 & 5 & 2.07512 & 2.25897E-02 & 2.07974 & 2.00124E-01 & 102 \\
0.5 & 10 & 2.07538 & 9.97694E-03 & 2.07738 & 8.63624E-02 & 301 \\
0.5 & 20 & 2.07563 & 2.04008E-03 & 2.07656 & 4.67255E-02 & 778 \\
0.5 & 50 & 2.07556 & 1.16593E-03 & 2.07595 & 1.74975E-02 & 3818 \\
\hline
1.0 & 5 & 2.13025 & 1.84896E-01 & 2.14786 & 6.40234E-01 & 104 \\
1.0 & 10 & 2.13232 & 8.78697E-02 & 2.13995 & 2.69313E-01 & 263 \\
1.0 & 20 & 2.13489 & 3.25920E-02 & 2.13750 & 1.54844E-01 & 761 \\
1.0 & 50 & 2.13440 & 9.35931E-03 & 2.13551 & 6.11910E-02 & 4604
\end{tabular}
\caption{Accuracy and runtime for the deep SMP method for various terminal times $T$ and numbers of time steps $N$ applied to the unconstrained power utility problem in a stochastic volatility model.}

\label{hest_SMP}
\end{table}
\end{example}

\begin{remark} We have applied this methodology to investing in the Vasicek model where the interest rate is a random process rather than the volatility. In the same way as for the Heston model we consider it both as a 2-dimensional state process and a 1-dimensional state process with random coefficients. We find similar results as for the Heston model, so we omit this example for brevity.
\end{remark}

\begin{example}[Heston with non-HARA utility]
We now look at a higher dimensional example with no explicit solution. Consider the same stochastic volatility market, but with multiple stocks. For simplicity we assume all stocks are independent with the same parameters, but this is not a necessary restriction. We let $U$ have the form (\ref{nonHARA}).  For this model we do not have a convenient ansatz for $u$ to obtain a solution, so we  simply compare our algorithms. Say we have $k$ stocks. We would then have control processes $\pi(t) \in \R^k$, $\Gamma_1(t) \in \R^{1+k,1+k}$ for the primal 2BSDE algorithm, $v(t) \in \R^{2k}$, $\Gamma_2(t) \in \R^{1+k,1+k}$ for the dual 2BSDE, and $v(t), Q_2(t) \in \R^k$ for the SMP algorithm. 

For this example let us take $T = 0.2$, $N = 20$, and $k = 10$. We then have $m = 10$, $d = 11$, and $n = 20$ in (\ref{update}). For the 2BSDE algorithms, the combined dimensions of the unknown processes are 131 for the primal and 141 for the dual. We consider the unconstrained problem, so we can set $\eta = 0$. For the SMP algorithm we only need to find processes in 20 dimensions. We run the algorithm for 10000 steps. Table \ref{stochastic_nonhara} shows the results of this. We give the value approximation for the 2BSDE algorithms, and the average of the upper and lower bound for the SMP. We also provide the runtime to show the efficiency of the algorithms as a whole, but the individual differences should be taken with a pinch of salt due to implementation differences, see Remark \ref{runtime}. The runtime has increased from the previous low dimensional example, but sub-linearly. The relative difference between approximations has increased to about 1\%.

\begin{table}[h]
\centering
\begin{tabular}{c||c|c}
Method & Value & Time (s) \\
 \hline  \hline
Primal & 2.61059 & 2316 \\
Dual & 2.63758 & 5636 \\
SMP & 2.62514 & 1840 \\
\end{tabular}
\caption{Value approximations for the 2BSDE and SMP algorithms applied to the unconstrained non-HARA utility problem in a stochastic volatility model.}
 \label{stochastic_nonhara}
\end{table}

\begin{remark} \label{runtime} There are a number of factors that affect the algorithm run times. For example, we see that the dual algorithm takes twice as long as the primal, but we did not simplify the algorithm to reflect the constraints of the problem. Making this change would have reduced the run time. Additionally, the $Z$ processes defined in (\ref{2BSDE}) involve a differentiation element. This differentiation is done `automatically', as it is handled by the computer, as opposed to the explicit derivative for the BSDE (\ref{utility_BSDE_dual}). This does not affect the accuracy, but increases the run time.
\end{remark}

\end{example}

\begin{example}[Power Utility in a Path Dependent Volatility Model]
In this example not only is the control problem non-Markovian, but it cannot be made Markovian by including additional information in our state space. This means that we cannot apply our 2BSDE method. 

This example is motivated by the concept that the volatility of a stock should be dependent on its value. If the stock is doing well the volatility is assumed to be lower. This may be due to infrequent trading of the stock, or the stability of the large company to which the stock is assigned. We consider an extreme example, where there are two levels of volatility. The normal higher level is maintain until the stock reaches its historical maximum, in which case it instantly transitions to the lower volatility.

Consider the market $(\ref{market})$, where the process $\sigma$ is a function of the stock history. For each $t \in [0,T]$, let $\sigma(t)$ be a diagonal matrix, with $i$th diagonal entry $\sigma_i(t)$ defined as
\[\sigma_i(t) = \begin{cases}
\sigma_\text{high}, & S_i(t) < \sup_{s \leq t}S_i(s), \\
\sigma_\text{low}, & S_i(t) = \sup_{s \leq t}S_i(s),
\end{cases}
\]
where $\sigma_\text{low} < \sigma_\text{high}$ are constants and $S_i$ is the $i$th stock price. In this model, the volatility of a stock is inversely proportional to the ratio of its value and the running maximum. This makes the diffusion coefficient of the wealth process $X$ a function of the entire path of $S$, instead of just at one time. This path dependence ensures that the utility maximisation problem is truly non-Markovian, in the sense that it cannot be made Markovian by the introduction of a new variable. We do not need to apply an artificial stock market argument here, as the volatility is a measurable function of the adapted stock market process up to time $t$, hence is progressively measurable.

For our application, we consider a $2$-dimensional stock market with cone constraint $K = \R^2_+$. The constraining function we use for this case is $h_K(x) = x^2$. We take $\sigma_\text{low} = 0.2$ and $\sigma_\text{high} = 0.3$. We then take $r(t) = 0.05$, $\mu(t) = 0.06\ind_2$ and $p = 0.5$. Table \ref{path} shows the duality gap when this algorithm is run for varying terminal times and numbers of time steps. We run the algorithms for 10000 iteration steps. The `Diff' and `Rel-Diff' columns show the absolute and relative distances between the upper and lower approximations respectively. We see in all cases that the duality gap reduces as the number of time steps increases. We do not have a benchmark to compare these results to, as the path dependence and incompleteness of the problem make it very difficult to solve analytically.

\begin{table}[h]
\centering
\begin{tabular}{c|c||c|c||c|c}
T & N & $u_\text{low}$ & $u_\text{high}$ & Diff & Rel-Diff (\%) \\
 \hline  \hline

0.2 & 5 & 2.01057 & 2.01066 & 8.99222E-05 & 4.47246E-03 \\
0.2 & 10 & 2.01054 & 2.01063 & 8.55058E-05 & 4.25287E-03 \\
0.2 & 20 & 2.01057 & 2.01063 & 6.00701E-05 & 2.98772E-03 \\
0.2 & 50 & 2.01061 & 2.01062 & 1.30984E-05 & 6.51462E-04 \\
\hline
0.5 & 5 & 2.02652 & 2.02682 & 3.03732E-04 & 1.49879E-02 \\
0.5 & 10 & 2.02654 & 2.02676 & 2.11323E-04 & 1.04278E-02 \\
0.5 & 20 & 2.02648 & 2.02665 & 1.73093E-04 & 8.54166E-03 \\
0.5 & 50 & 2.02637 & 2.02650 & 1.31536E-04 & 6.49120E-03 \\
\hline
1.0 & 5 & 2.05333 & 2.05421 & 8.87694E-04 & 4.32320E-02 \\
1.0 & 10 & 2.05337 & 2.05397 & 5.93055E-04 & 2.88821E-02 \\
1.0 & 20 & 2.05327 & 2.05368 & 4.16821E-04 & 2.03005E-02 \\
1.0 & 50 & 2.05307 & 2.05339 & 3.20542E-04 & 1.56128E-02
\end{tabular}
\caption{Duality gap for the deep SMP method for various terminal timed $T$ and numbers of time steps $N$ applied to the cone-constrained power utility problem with path dependent volatility.}
 \label{path}
\end{table}

\end{example}

\section{Methodology Comparison} \label{method}

First we compare the results of small changes to the methodology of the DC2BSDE algorithm,  then compare the algorithms we have introduced to others existing in the literature.

\subsection{Parameter Comparison} \label{parameter}

 We consider the primal deep controlled 2BSDE algorithm for the unconstrained Merton problem, with $T = 1.0$ and $m  = 10$ unless stated otherwise, and the other coefficients chosen as in Example \ref{HARA}. These tests were performed on an HP ZBook Studio G4 with a 2.9 GHz Intel i7 CPU, with no GPU acceleration.

The first comparison we consider is the activation function used within the neural networks (c.f. function $h$ in the description of the algorithm). Table \ref{compare_act} shows the results of this test, where we take $N = 20$. The choice of activation function is not important for this particular problem. All the functions are valid for this algorithm, since they are continuous functions that are not polynomials, see \cite{leshno1993multilayer}. 
\begin{table}[h]
\centering
\begin{tabular}{c|c||c|c}
Name & Function  & Rel - Err (\%) & Time (s) \\
 \hline  \hline
ReLu & $x \mapsto \max(0,x)$ & 3.57771E-03 & 352.7 \\
softplus & $x \mapsto \log(1 + e^x)$  &  3.57250E-03 & 352.0 \\
tanh & $x \mapsto \tanh(x)$ & 3.44745E-03 & 352.8 \\
sigmoid & $x \mapsto \frac{e^x}{e^x + 1}$ & 3.53815E-03 & 352.0
\end{tabular}
\caption{Accuracy and runtime of the primal deep controlled 2BSDE method using different activation functions, applied to the unconstrained power utility problem.} 
\label{compare_act}
\end{table}

The second comparison we make is the optimisation method we use to minimise the loss function. We compare ADAM to standard gradient descent, the momentum optimiser (with momentum parameter 0.9), and the ADAGRAD optimiser, which are described in \cite{ruder2016overview}. Table \ref{compare_opt} shows the results of this test. ADAM is the best optimiser for this case and suggests also why it is the most popular in the literature. However, there is some time loss for using this method, compared to the other optimisers. A reason for the much higher convergence rate for the ADAM algorithm may be that it is better suited to a higher learning rate. Therefore, to get the same convergence results using the other optimisation algorithms one may need to run the other algorithms for longer, with a lower rate, which would cancel out the time loss.

\begin{table}[h]
\centering
\begin{tabular}{c||c|c}
Optimiser & Rel - Err (\%) & Runtime (s) \\
 \hline  \hline
Gradient Descent & 1.00474E-01 & 358.3 \\
ADAM & 3.54173E-03 & 384.6 \\
ADAGRAD & 6.84528E-02 & 334.8 \\
Momentum & 5.16220E-02 & 338.1
\end{tabular}
\caption{Accuracy and runtime of the primal deep controlled 2BSDE method using different optimisation algorithms, applied to the unconstrained power utility problem.} 
\label{compare_opt}
\end{table}

\begin{figure}[h]
\centering
\begin{minipage}{.5\textwidth}
\centering
\begin{subfigure}[b]{\textwidth}
\includegraphics[width=\textwidth]{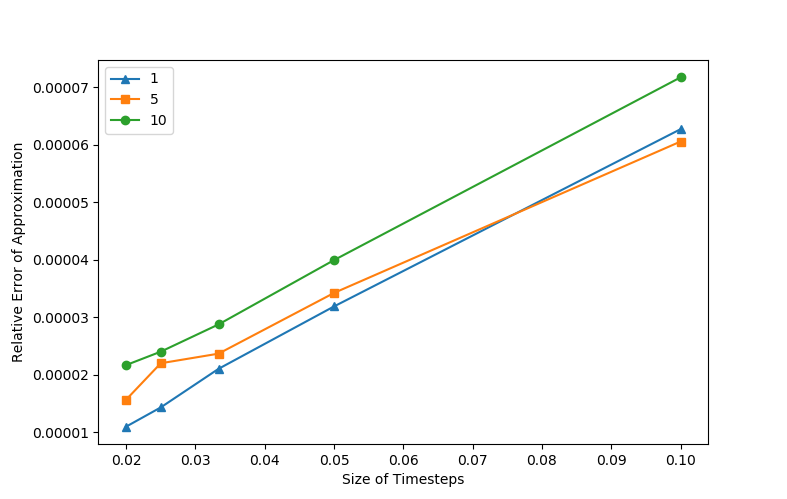}
\caption{Approximation Error of Methods.}
\end{subfigure}
\end{minipage}%
\begin{minipage}{.5\textwidth}
\centering
\centering
\begin{subfigure}[b]{\textwidth}
\includegraphics[width=\textwidth]{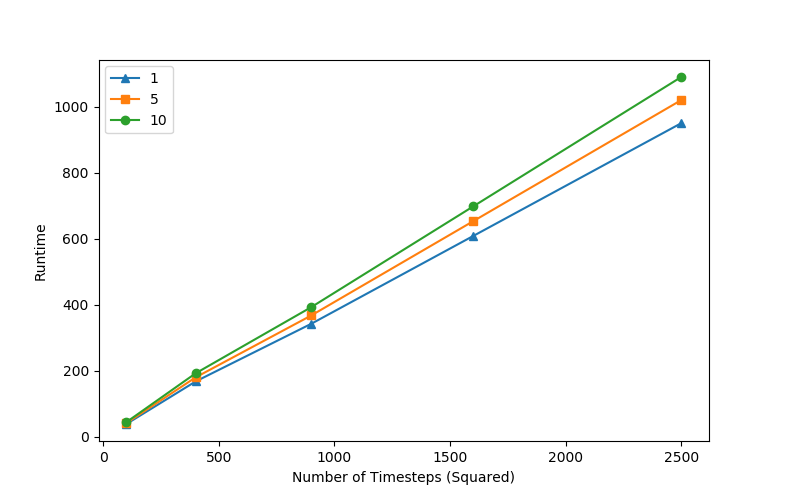}
\caption{Runtime of Methods.}
\end{subfigure}
\end{minipage}
\caption{Accuracy and runtime of the primal deep controlled 2BSDE method applied to unconstrained power utility problems with different dimensions.} 
\label{compare_dim}
\end{figure}

Now we compare the running time and relative error given changes in the dimensionality of the problem. We compare dimensions $m \in \{ 1 , 5, 10\}$ of the stock market, with numbers of time steps $N \in \{10,20,30,40,50\}$. Figure \ref{compare_dim} shows the results of this test. The first graph shows the error of approximation against $N^{-1}$, and the second shows runtime against $N^2$. We see the order $\ord(N^{-1})$ convergence for each dimension, but the error increases with dimension. Therefore, to achieve the same accuracy, a larger number of time steps would be needed for a higher dimension. The dimensionality complexity of this algorithm is hence tied to the complexity with respect to the number of time steps $N$, which looks from the graph to be of order $\ord(N^2)$. The time difference between the three dimensions is minimal. However, this is for a set number of iteration steps, and for even higher dimensions more steps may be needed, which would increase the runtime compared to lower dimensional models.

\begin{figure}[h]
\centering
\begin{minipage}{.5\textwidth}
\centering
\begin{subfigure}[b]{\textwidth}
\includegraphics[width=\textwidth]{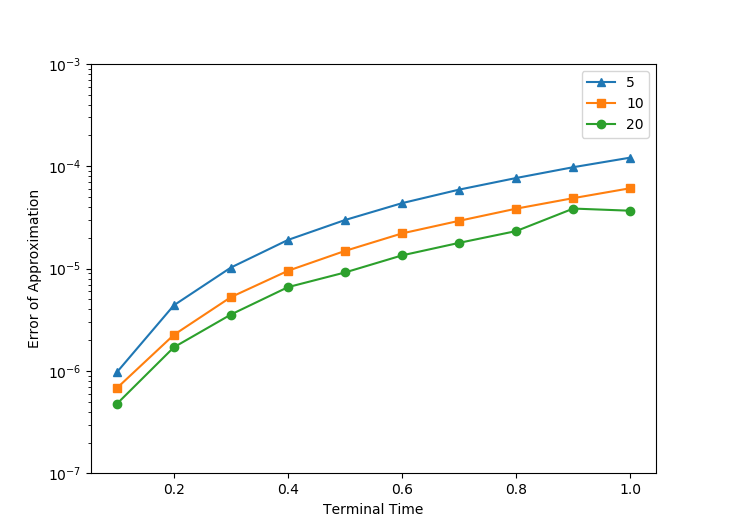}
\caption{Log-Scaled Approximation Error.}
\end{subfigure}
\end{minipage}%
\begin{minipage}{.5\textwidth}
\centering
\centering
\begin{subfigure}[b]{\textwidth}
\includegraphics[width=\textwidth]{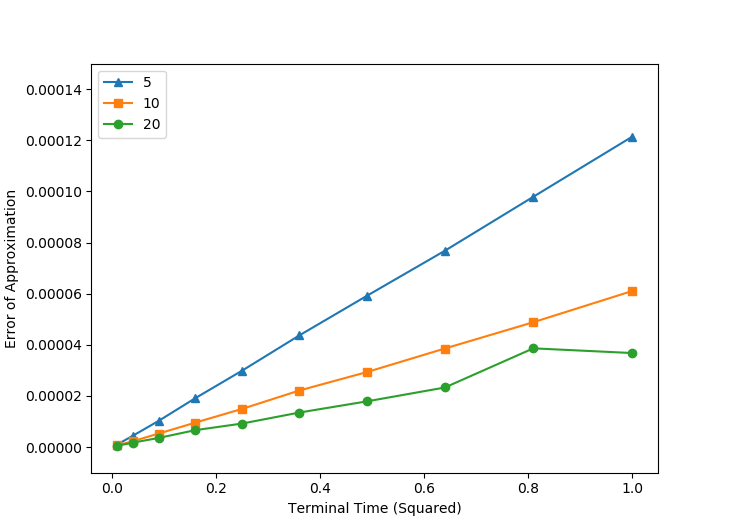}
\caption{Square-Scaled Approximation Error.}
\end{subfigure}
\end{minipage}
\caption{Accuracy of the primal deep controlled 2BSDE method using different terminal times and numbers of time steps, applied to the unconstrained power utility problem. Both graphs display the same data, but the left uses a log scale for the y axis, and the right a squared scale for the x axis.} 
\label{compare_time}
\end{figure}

Finally, we consider the effect of the terminal time $T$ on the relative error of the output. Since nothing is changing in the algorithm, we have found that changing $T$ while keeping the number of time steps $N$ fixed does not change the runtime of the algorithm. However, Figure \ref{compare_time} shows an increasing error with $T$, as it increases from 0.2 to 1.0. Both graphs show the same data, the left side using a log scale for the y axis, and the right side squaring the x axis, to show the order $\ord(T^2)$ error implied by the data. We display this data using multiple time steps, from 5 to 20. It is notable here that even increasing the number of time steps by the same factor does not compensate for the error caused by increasing the terminal time. For example, the relative error when we use $(T,N) = (0.2,5)$ is 0.05\%, whereas when we use $(T,N) = (0.4,10)$ it is 0.12\%, which is just over double the error. This $\ord(T^2)$ vs $\ord(N^{-1})$ error may be why an assumption of $T$ being `small' is used in the convergence analysis of the first order BSDE solver, see \cite{han2018convergence}. 

{ 
Finally we consider the impact of the changes discussed in Remarks \ref{one_net} and \ref{one_loss} for a unconstrained power utility problem with $T = 0.5$, $m = 10$ and $N = 20$. The results are shown in Table \ref{compare_net}. We give the run time and accuracy for our method  (the first row) and the two alternatives. The second row compares the accuracy and run time of using a single neural network to define the control and unknown BSDE processes, against using a separate one for each time step. We see that the error has slightly increased, but mainly the runtime increases when we use a single net. This is due to the additional time needed to take derivatives on the neural network parameters, even though there are less of then. The last row compares the accuracy and runtime if we optimise both networks using a single loss function
\begin{align*}
\LL(\Theta^0_{\text{BSDE}},(\theta^0_i)_{i = 0}^{N-1}) =  \LL_1(\Theta^0_{\text{BSDE}}) + \sum_{i=0}^{N-1}\LL_2(\theta^0_i,i),
\end{align*}
which is a linear combination of the original two loss functions (\ref{L_1}) and (\ref{L_2}). We see a much higher runtime and a lower accuracy. 
%It is notable that even a somewhat accurate result is found in this situation since there is no theoretical basis for using this loss function.
}

\begin{table}[h] 
\centering
{ 
\begin{tabular}{c||c|c}
Method       & Rel-Err (\%) & Time (s) \\
\hline \hline
Normal       & 6.31965E-03   & 334      \\
\hline
``One Net''  & 1.93321E-02   & 1080      \\
\hline
``One Loss'' & 3.44806E-01   & 1689    
\end{tabular}
}
\caption{
{  Relative error and runtime of variations of the primal DC2BSDE algorithm for the unconstrained Power Utility problem}}
\label{compare_net}
\end{table}

\subsection{Comparison to Literature}

Now we compare the performance of our algorithm to two existing in the literature. We first compare to the algorithm of \cite{han2016deep} where the primal gains function is optimised directly. This is not explicitly given a name by the authors, so we refer to it as the PVO (Primal Value Optimisation) algorithm. Secondly we compare to the Hybrid-Now algorithm of \cite{hure2018deep1} (referred to in the sequel as Hybrid), where the control is optimised using an approximation of the value function, which is also found using neural networks. First, we note that while both algorithms search for a Markovian control, and as such solve Markovian control problems, they are potentially less restrictive than our algorithm. The maximum requirement for these algorithms is satisfaction of the dynamic programming principle, as opposed to our derivation of the algorithm relying on the HJB equation. However it may be possible that our algorithm works in such a general setting as well (and resulting approximate solutions to the Hamiltonian PDE become viscosity solutions). The second algorithm is formalised in terms of successive approximations of the value function working backwards from terminal time in a discretisation. This method requires suitable knowledge of the distribution of the optimal state process at each time step, which we do not know a priori. Instead, we  implement the algorithm in a similar vein to our own, generating sample paths of the state and control process and solving each optimisation problem \cite[(2.1)]{hure2018deep2} simultaneously. We  run each algorithm for a fixed number of steps of gradient descent. This may not be the best way to compare the algorithms since they may converge after a different number of steps, but in practise it is difficult to identify exactly when this happens. We only compare to the primal value approximations of our DC2BSDE and DSMP methods, though it would of course be possible to use PVO and Hybrid to solve the dual problem.

\begin{example}[Power Utility in Black Scholes Model]

Consider the utility maximisation problem as of Example \ref{HARA}, but we take the power ($p=\half$) utility $U(x) = 2\sqrt{x}$. We fix the terminal time $T = 0.5$ but change the stock market dimension $m$ and the number of time steps $N$. The other parameters remain the same as in Example \ref{HARA}. Table \ref{compare_lit} shows the results of this comparison. The solutions can be found (as is found in Example \ref{cone} but with no constraints) to be 2.04864 and 2.09756 for dimensions 5 and 20 respectively. For our methods we ran the DC2BSDE with $m = 20$ for 10000 steps, and 5000 steps for the $m = 5$ case and the DSMP.

 The comparison to the PVO algorithm indicates that our method sacrifices slight accuracy, but scales much better with number of time steps and dimension. Indeed, when $m = 20$ the lower accuracy can be compensated by doubling the number of time steps without reaching the runtime of the PVO algorithm. We ran the PVO algorithm with a learning rate of 0.01 for 10000 and 20000 steps for $m = 5, 20$ respectively. The results for the Hybrid algorithm originate from our own implementation of the method. There may be a number of differences from the original implementation which could lead to less accurate results than expected. For example, we are not sure how exactly the authors used the `explore first, exploit later' technique \cite[Section~3.3]{hure2018deep1} to generate sample paths of the optimal state process without first determining a control. In \cite{hure2018deep2}, a 100 dimensional PDE (that is, $d = 100$ using the notation of Section \ref{Markovian}) is solved with an accuracy of order 0.1\% and a runtime of 2 - 4 hours, using $N = 40$ time steps and a terminal time of $T = 1.0$. This is much better than our preliminary results for this algorithm. However, we did not see any convergence like this in our implementation, in any time frame that was comparable to our methods. The hybrid algorithm is the fastest per iteration step, however we ran the algorithm for 50000 steps with a learning rate of 0.0001 to avoid instability issues, which is the cause of the large runtime. 

\begin{table}[h]
\centering
\begin{tabular}{c|c||c|c|c||c|c|c}
$m$  & $N$  & DC2BSDE Value & Rel-Err (\%) & Time (s) & DSMP  Value & Rel-Err (\%) & Time (s) \\ \hline \hline
5  & 5  & 2.0483   & 1.4829E-02   & 54          & 2.0465      & 1.0476E-01 & 59   \\ 
5  & 10 & 2.0485   & 9.2087E-03   & 128         & 2.0505       & 9.0012E-02 & 148  \\
5  & 20 & 2.0486   & 4.4761E-03   & 309         & 2.0476      & 5.3657E-02 & 358  \\ \hline 
20 & 5  & 2.0942   & 1.6192E-01   &  92       & 2.0906       & 3.3063E-01 & 66   \\ 
20 & 10 & 2.0958  &  8.2069E-02  & 223         & 2.0947      & 1.3682E-01 & 167  \\ 
20 & 20 & 2.0961   &  6.5495E-02 &   538       & 2.0963     & 5.8714E-02 & 412  \\  \hline  \hline 
 $m$  & $N$   & PVO Value   &  Rel-Err (\%) & Time (s)     & Hybrid Value  &   Rel-Err (\%) & Time (s)\\  \hline  \hline 
5  & 5  & 2.0497   & 4.9400E-02   & 79          & 2.0441    & 2.2415E-01 & 328  \\
5  & 10 & 2.0478  & 4.3425E-02   & 266         & 2.0457      & 1.4347E-01 & 479  \\
5  & 20 & 2.0483  & 1.8711E-02   & 950         & 2.0467      & 9.5190E-02 & 1630 \\ \hline 
20 & 5  & 2.0952   & 1.1184E-01   & 233         & 2.0431      & 2.5977E+00 & 343  \\
20 & 10 & 2.0963   & 5.9899E-02   & 813         & 2.0536      & 2.0940E+00 & 813  \\
20 & 20 & 2.0969   & 3.0629E-02   & 1824        & 2.0604      & 1.7720E+00 & 1747
\end{tabular}
\caption{Comparison of the Primal DC2BSDE and (primal value approximation of the) DSMP algorithm to the PVO and Hybrid algorithms for the unconstrained Power Utility problem.}
\label{compare_lit}
\end{table}

{
The example we consider is a toy example with an analytical solution. In particular, the supremum term inside the HJB equation (\ref{HJB}) can be written exactly and for this problem we are left with the nonlinear PDE

\[\frac{d}{dt}u(t,x) = \half |\theta(t)|^2 \frac{D_xu(t,x)^2}{D_{xx}u(t,x)} - rxD_xu(t,x).\]

This PDE is control independent. In particular it is independent of the control dimension $m$ apart from the negligible time taken to calculate $\theta(t)$. The PDE can be solved directly using the Deep 2BSDE method introduced in \cite{beck2017machine}. The results of solving this for $m = 20$ are given in Table \ref{PDE_method}. Due to the second derivative in the denominator, this PDE is unstable. In the algorithm we require 50000 iteration steps to see convergence results. Therefore, while this algorithm is faster per iteration step, it is slower overall. However its runtime would be comparable to our algorithm for larger value of $m$. It is also an order of magnitude less accurate, and can only be used in the special case where the supremum term can be written out as an explicit function of $t$ and $x$. 

The primal HJB equation, even in this simple case, can be very difficult to solve numerically. This further emphasises the utility of the algorithms presented in this paper, solving the equation by either finding the control without nonlinear dependence on the value function derivatives, or by solving the dual problem.
}

\begin{table}[h]
\centering
{
\begin{tabular}{c|c|c|c}
N  & Value       & Rel-Err(\%) & Time (s) \\
\hline \hline
5  & 1.96570E+00 & 6.28613E+00 & 363      \\ \hline
10 & 2.05429E+00 & 2.06281E+00 & 865      \\ \hline
20 & 2.06381E+00 & 1.60905E+00 & 1980    
\end{tabular}
}
\caption{
{ Value, error and runtime of the Deep 2BSDE method for the unconstrained Power Utility problem.}
\label{PDE_method}}
\end{table}

\end{example}

\section{Conclusion}

In this paper we study a variety of constrained utility maximisation problems, including some that are incomplete or non-Markovian, and introduced two algorithms, called the deep controlled 2BSDE method and the deep SMP method, for solving such problems, which involve simulating a system of BSDEs of first or second order. We use these equations, together with the optimality conditions for the optimal control, to find the value function of the problem. We apply these algorithms to a range of utility functions and closed convex control constraints, and find highly accurate solutions, with order $\ord(N^{-1})$ convergence in terms of the number of time steps of the discretisation. We derive tight bounds of the value function at time 0 using the convex duality method that results in a primal problem and a dual problem, both can be solved with our algorithms.
There remain many interesting open questions such as  convergence analysis,  duality theory  for multidimensional
 state processes,  the efficiency of the algorithm, etc.   We leave these questions and others for future research.

{ \bigskip\noindent
{\bf Acknowledgments}. The authors are grateful to  two anonymous reviewers whose constructive comments and suggestions have helped to improve the paper of the previous version.
}

\appendix
\section*{Appendix}

\section{Proof of Theorem \ref{2BSDE_thm}}

\textit{Proof of \ref{2BSDE_thm}.}
Let $\alpha$ denote the optimal control process, and $\mathcal{X}$ the corresponding state process. To make notation simpler, without loss of generality let $d = 1$, so the state space is one dimensional, and we let $\Sigma$ be a vector in $\R^n$, rather than a matrix in $\R^{1 \times n}$. Define new processes on $[0,T]$ by
\begin{align}
 V(t) := u\left(t,\mathcal{X}(t)\right), \quad 
Z(t) := \frac{\partial}{\partial x}u\left(t,\mathcal{X}(t)\right), \quad
\Gamma(t) := \frac{\partial^2}{\partial x^2} u\left(t,\mathcal{X}(t)\right).
\label{value_solutions}
\end{align}
By the regularity assumptions on $u$, these processes are continuous. Furthermore by construction (\ref{maxHam}) is satisfied. Applying It\^o's Lemma to $u$ and using the HJB equation, we get

\begin{align*}
dV(t) &= Z(t) d\mathcal{X}(t) + \half \Gamma(t)d\langle \mathcal{X},\mathcal{X}\rangle (t) + \frac{\partial }{\partial t}u(t,\mathcal{X}(t))dt\\
& = \left[ F\left( t, \mathcal{X}(t),\alpha(t),Z(t),\Gamma(t)\right) - \sup_{a \in A} F\left(t,\mathcal{X}(t),a,Z(t),\Gamma(t)\right) \right] dt + Z(t)\Sigma\left(t,\mathcal{X}(t),\alpha(t)\right)^\top dW(t), \\
&= Z(t)\Sigma\left(t,\mathcal{X}(t),\alpha(t)\right)^\top dW(t),
\end{align*}
with the terminal condition $V(T) = g(\mathcal{X}(T))$. We assume $u$ is three times continuously differentiable in $x$, so we can similarly apply It\^o's lemma to  $\frac{\partial}{\partial x}u$. Then
\begin{align*}
dZ(t) &= \Gamma(t)^\top d\mathcal{X}(t) + \half \frac{\partial^3}{\partial x^3} u(t,\mathcal{X}(t)))   d\langle \mathcal{X},\mathcal{X}\rangle (t) + \frac{\partial^2}{\partial x \partial t}u(t,\mathcal{X}(t))dt
\end{align*}
with terminal condition is $Z(T) = g'(\mathcal{X}(T))$, where
\begin{align*}
&\frac{\partial^2}{\partial x\partial t}u(t,\mathcal{X}(t)) = -\frac{\partial}{\partial x} \sup_a F(t, \mathcal{X}(t), a, Z(t), \Gamma(t)) \\
& = -\frac{\partial}{\partial x} \left[b(t,\mathcal{X}(t), \alpha(t))^\top \frac{\partial}{\partial x} u(t,\mathcal{X}(t)) + \half \tr \left( \frac{\partial^2}{\partial x^2}u(t,\mathcal{X})\Sigma(t, \mathcal{X}, \alpha(t))^\top  \Sigma(t, \mathcal{X}(t), \alpha(t))\right)\right] \\
& = - (\frac{\partial}{\partial x} b(t,\mathcal{X}(t), \alpha(t))\frac{\partial}{\partial x} u(t,\mathcal{X}(t))) - b(t,\mathcal{X}(t), \alpha(t)) \frac{\partial^2}{\partial x^2} u(t,\mathcal{X}(t)) \\
& - \frac{\partial^2}{\partial x^2}u(t,\mathcal{X}(t)) \Sigma(t,\mathcal{X}(t),\alpha(t))^\top  \frac{\partial}{\partial x} \Sigma(t, \mathcal{X}(t), \alpha(t)) - \half |\Sigma(t, \mathcal{X}(t), \alpha(t))|^2 \frac{\partial^3}{\partial x^3}u(t,\mathcal{X}(t))
\end{align*}
Cancellation then yields 
\begin{align*}
dZ(t) & = -\left(Z(t)\frac{\partial}{\partial x}b(t, \mathcal{X}(t), \alpha(t)) + \Gamma(t)\Sigma(t,\mathcal{X}(t), \alpha(t))^\top \frac{\partial}{\partial x} \Sigma(t, \mathcal{X}(t), \alpha(t))\right) dt \\
&+ \Gamma(t) \Sigma(t, \mathcal{X}(t), \alpha(t))^\top dW(t)
\end{align*}
as required. \qed

\section{Algorithm}

\begin{algorithm}[H] \label{primal_DC2BSDE} 
\KwIn{$x \in \R^d$}
\SetAlgoLined
\KwResult{$u(0,x)$ }
Initialise $V^0_0 \sim \text{Unif}([-0.1,0.1])$ and $Z^0_0 \sim \text{Unif}([-0.1,0.1]^{d})$\;
 \For{$i = 0,1 ,2 ,\ldots, N-1$}{ 
 Initialise $\theta^0_i\sim \text{Unif}([-0.1,0.1]^{\rho(d,m)})$, $\lambda^0_i \sim \text{Unif}([-0.1,0.1]^{\rho(d, d^2)})$\;
 }
 Set $\Theta^0_1 = \{V^0_0,Z^0_0\} \cup \{\lambda^0_i\}_{i=0}^{N-1}$\;
 Set $\Theta^0_2 = \{\theta_i^0\}_{i=0}^{N-1} $\;
 
 Initialise $m_0 = v_0 = 0, j = 0$ and $m_0^i = v_0^i = 0$ for $i = 0, \ldots, N-1$\;
\While{$\Theta^j_1, \Theta^j_2$ not converged}{
	Generate $W_k$ for $k = 1,\ldots B$\;
	Generate $\mathcal{X}^k,V^k,Z^k, \alpha^k, \Gamma^k$ using (\ref{2BSDE_disc}) and $\left(\Theta^{j}_1, \Theta^{j}_2\right)$\;
	Set $L^j_1(\Theta) = \frac{1}{B}\sum_{k=1}^B \left| V^k_N - g(\mathcal{X}^k_N)\right|^2 + \half \left|Z^k_N - D_x g(\mathcal{X}^k_N)\right|^2$\;
	Set $\Theta^{j+1}_1, m^1_{j+1}, v^1_{j+1} = $ ADAM($\Theta^1_{j}, L^1_{j}, m^1_{j}, v^1_{j}, j$)\;
	Regenerate $\mathcal{X}^k,V^k,Z^k, \alpha^k, \Gamma^k$ using (\ref{2BSDE_disc}) and $\left(\Theta^{j+1}_1, \Theta^{j}_2\right)$\;
	 \For{$i = 0,1 ,2 ,\ldots, N-1$}{
	 Set $L^j_2(\theta, i) = -\frac{1}{B}\sum_{k=1}^B 	F( t_i, \mathcal{X}^k_i,\alpha^k_i, Z^k_i, \Gamma^k_i)	$\;
	Set $\theta^{j+1}_i, m^i_{j+1}, v^i_{j+1} = $ ADAM($\theta^{j+1}_i, L^j_{2}(\cdot, i), m^i_{j}, v^i_{j}, j$)\;
	}
	 Set $\Theta^{j+1}_2 = \{ \theta^{j+1}_i\}_{i=0}^{N-1}$\;
	Set $j = j+1$\;
	}

 Output $V_0^j$\;
  \caption{Primal Deep Controlled 2BSDE (for maximisation)}
\end{algorithm}

\bibliographystyle{abbrv}
\bibliography{Biblio}
\end{document}